%% file: main.tex
\documentclass[letterpaper,twocolumn,10pt]{article}

\usepackage{xr-hyper}
\makeatletter
\newcommand*{\addFileDependency}[1]{
\typeout{(#1)}
\@addtofilelist{#1}
\IfFileExists{#1}{}{\typeout{No file #1.}}
}
\makeatother
\newcommand*{\myexternaldocument}[1]{%
 \externaldocument{#1}%
 \addFileDependency{#1.tex}%
 \addFileDependency{#1.aux}%
}

\usepackage{usenix-2020-09}
\usepackage[all=normal,paragraphs=tight,floats=tight,wordspacing=tight,tracking=tight,bibbreaks=tight]{savetrees}
\usepackage[compact,small]{titlesec}
\usepackage{xspace}
\usepackage{outlines}
\usepackage{siunitx}
\usepackage{xcolor}
\usepackage{subcaption, newfloat, wrapfig}
\usepackage[labelfont={small,it,bf},textfont={small,it}]{caption}
\usepackage{booktabs}
\usepackage{amsmath, amsthm, mathtools}
\usepackage{xurl}
\usepackage[frozencache=true,cachedir=minted-cache]{minted}
\usepackage{mdframed}
\usepackage{dashrule}
\usepackage{paralist}
\usepackage{enumitem}
\usepackage{cleveref}
\usepackage{listings}
\usepackage{pifont}
\usepackage{braket}
\usepackage{threeparttable}

\usepackage{tikz}
\usetikzlibrary{calc,tikzmark,arrows.meta,positioning}

\usepackage{algorithm}
\usepackage[noend]{algpseudocode}

\usepackage[available,functional,reproduced]{usenixbadges}

\definecolor{tol1}{HTML}{332288}
\definecolor{tol2}{HTML}{117733}
\definecolor{tol3}{HTML}{44AA99}
\definecolor{tol4}{HTML}{88CCEE}
\definecolor{tol5}{HTML}{DDCC77}
\definecolor{tol6}{HTML}{CC6677}
\definecolor{tol7}{HTML}{AA4499}
\definecolor{tol8}{HTML}{882255}

\newif\ifextended
\extendedtrue

\ifextended
    \newcommand{\refAppendix}[1]{\Cref{#1}}
\else
    \myexternaldocument{appendixStandalone}
    \newcommand{\refAppendix}[1]{our extended paper~\cite[\S~\ref*{#1}]{extended}}
\fi

\lstnewenvironment{compactsql}{\lstset{language=sql,aboveskip=0pt,belowskip=0pt,xleftmargin=0em}}{}

\usepackage{pdfcomment}
\newenvironment{mychange}{\begin{pdfsidelinecomment}[color=gray,icolor=yellow,caption=inline,linebegin={/Butt},lineend={/Square},linewidth=2bp,linesep=.5cm]{Changed}}{\end{pdfsidelinecomment}}
\renewenvironment{mychange}{}{}

\algnewcommand\algorithmicswitch{\textbf{switch}}
\algnewcommand\algorithmiccase{\textbf{case}}
\algdef{SE}[SWITCH]{Switch}{EndSwitch}[1]{\algorithmicswitch\ #1\ \algorithmicdo}{\algorithmicend\ \algorithmicswitch}%
\algdef{SE}[CASE]{Case}{EndCase}[1]{\algorithmiccase\ #1\textbf{:}}{\algorithmicend\ \algorithmiccase}%
\algtext*{EndSwitch}%
\algtext*{EndCase}%

\newcommand{\SIIntPercent}[1]{\SI[round-mode=places,round-precision=0]{#1}{\percent}}

\DeclareFloatingEnvironment[
  fileext=lop,
  listname={List of Listings},
  name=Listing
]{mylisting}
\DeclareCaptionSubType{mylisting}
\crefname{mylisting}{listing}{listings}
\Crefname{mylisting}{Listing}{Listings}
\crefname{submylisting}{listing}{listings}
\Crefname{submylisting}{Listing}{Listings}

\crefname{section}{\S}{\S\S}
\crefformat{section}{\S#2#1#3}

\mdfsetup{skipbelow=0}

\newcommand{\allnotes}[1]{}

\newcommand{\name}{Blockaid\xspace}
\newcommand{\diaspora}{diaspora*\xspace}
\newcommand{\marco}{\textsf{MARCO}\xspace}
\newcommand{\cvc}{\textsc{cvc5}\xspace}

\newcommand{\V}{\mathcal{V}}

\newcommand{\sfx}{\mathsf{x}}
\newcommand{\sfq}{\mathsf{Q}}
\newcommand{\sfv}{\mathsf{V}}
\newcommand{\T}{\mathcal{T}}
\newcommand{\D}{\mathcal{D}}
\newcommand{\bfx}{\mathbf{x}}
\newcommand{\bfc}{\mathbf{c}}
\newcommand{\bfy}{\mathbf{y}}

\newcommand{\req}{\mathit{req}}
\newcommand{\ctx}{\mathit{ctx}}

\newcommand{\Prog}{\mathcal{P}} 
\DeclareMathOperator{\NI}{NI}
\newcommand{\EnforcePred}{E}

\newcommand{\Tmin}{\T_{\text{min}}}
\newcommand{\Csmall}{C_{\text{small}}}
\newcommand{\Ccore}{C_{\text{core}}}
\newcommand{\Caug}{C_{\text{aug}}}

\newcommand{\cmark}{\text{\ding{51}}}%
\newcommand{\xmark}{\text{\ding{55}}}%

\newcommand{\paramd}[1]{#1^{\text{p}}}

\newcommand\eat[1]{}

\newcommand{\diffBox}[1]{\colorbox{black!15}{#1}}
\newcommand{\prm}[1]{\colorbox{black!15}{\texttt{#1}}}
 
\usepackage[small, compact]{titlesec}
\titlespacing{\paragraph}{%
  0pt}{
  0.1\baselineskip}{
  1em}

\setminted{escapeinside=||}

\newcommand{\sql}[1]{\mintinline{sql}{#1}}

\newlist{rrlist}{itemize}{1}
\setlist[rrlist]{label=$\hookrightarrow$,topsep=-.75em}

\newlist{qlist}{itemize}{1}
\setlist[qlist]{label={\includegraphics[width=.8em]{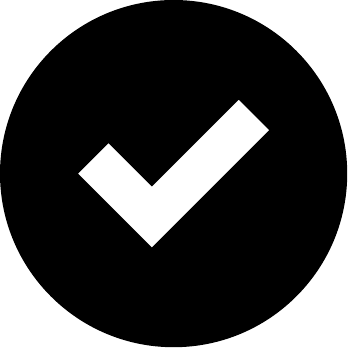}},topsep=-.75em}

\theoremstyle{plain}
\newtheorem{theorem}{Theorem}[section]

\newtheorem{proposition}[theorem]{Proposition}

\theoremstyle{definition}
\newtheorem{definition}[theorem]{Definition}
\newtheorem{example}[theorem]{Example}
\theoremstyle{plain}

\theoremstyle{remark}

\theoremstyle{plain}

\setlist{leftmargin=*,itemsep=0pt,parsep=0pt,topsep=1pt,partopsep=1pt}

\title{
    \Large \bf \name: Data Access Policy Enforcement for Web Applications
    \ifextended
    \\ (Extended Technical Report)
    \fi
}
\date{}

\usepackage{authblk}

\author{Wen Zhang$^{\text{1}}$}
\author{Eric Sheng$^{\text{2,*}}$}
\author{Michael Chang$^{\text{1}}$}
\author{Aurojit Panda$^{\text{3}}$}
\author{Mooly Sagiv$^{\text{4}}$}
\author{Scott Shenker$^{\text{1,5}}$}
\affil{
  $^1$UC Berkeley \hspace{0.2in}
  $^2$Yugabyte \hspace{0.2in}
  $^3$NYU \hspace{0.2in}
  $^4$Tel Aviv University \hspace{0.2in}
  $^5$ICSI
}

\begin{document}
\maketitle
{\renewcommand{\thefootnote}{*}\footnotetext{Work done while at UC Berkeley.}}

\begin{abstract}
\input{nbstract}
\end{abstract}

\section{Introduction}\label{sec:intro}
\input{intro}

\section{Related Work}\label{sec:related}
\input{related}

\section{System Design}\label{sec:design}
\input{design}

\section{View-based Policy and Compliance}\label{sec:spec}
\input{specification}

\section{Compliance Checking with SMT}\label{sec:checking}
\input{checking}

\section{Decision Generalization and Caching}\label{sec:cache}
\input{caching}

\section{Implementation}\label{sec:impl}
\input{impl}

\section{Evaluation}\label{sec:eval}
\input{eval}

\section{Additional Issues}\label{sec:discussion}
\input{discussions}

\section{Conclusion}
\input{conclusion}

\section*{Acknowledgments}
We are grateful to Alin Deutsch and Victor Vianu for the many discussions about query determinacy,
and to Nikolaj Bjørner, Alvin Cheung, Vivian Fang, and members of the Berkeley NetSys Lab for their help with the project.
We also thank the anonymous reviewers and our shepherd Malte Schwarzkopf for their helpful comments.
This research was funded in part by NSF grants 1817116 and 2145471, and gifts from Intel and VMware.

\bibliographystyle{plain}
\bibliography{blockaid}

\appendix
\input{artifactAppendix}

\ifextended
  \input{extendedAppendix}
\fi

\end{document}

%% file: nbstract.tex
Modern web applications serve large amounts of sensitive user data, access to which is typically governed by data-access policies.
Enforcing such policies is crucial to preventing improper data access, and prior work has proposed many enforcement mechanisms.
However, these prior methods either alter application semantics or require adopting a new programming model;
the former can result in unexpected application behavior, while the latter cannot be used with existing web frameworks.

\name{} is an access-policy enforcement system that preserves application semantics and is compatible with existing web frameworks.
It intercepts database queries from the application, attempts to verify that each query is policy-compliant, and blocks queries that are not.
It verifies policy compliance using SMT solvers and generalizes and caches previous compliance decisions for better performance.
We show that \name{} supports existing web applications while requiring minimal code changes and adding only modest overheads.

%% file: intro.tex
Many modern web applications use relational databases to store sensitive user data, access to which is governed by organizational or regulatory \emph{data-access policies}.
To enforce these policies, today's web developers wrap each database query within access checks that determine whether a user has access to the queried data.
As an application can query the database at many call sites, getting access checks right at every call site is challenging, and erroneous or missing checks have exposed sensitive data in many production systems%
~\cite{stock18:hotcrp_hidden_tags,kohler13:hotcrp_hide_rounds,kohler15:hotcrp_download_pc,ashford15:facebook_photo,maunder16:wordpress,green17:piazza}.

Prior work has suggested a variety of languages, frameworks, and tools that simplify the enforcement of data-access policies.
As we detail in \Cref{sec:related}, these approaches either
\begin{inparaenum}[(1)]
    \item require applications be written using specialized web frameworks, hindering their adoption; or
    \item transparently remove from query results any data that cannot be revealed, possibly resulting in unexpected application behavior (e.g., the user has no idea that there are missing results and reaches the wrong conclusion).
\end{inparaenum}

This paper proposes an alternative approach to enforcing data-access policies that meets four goals:
\begin{enumerate}
    \item \textbf{Backwards compatibility}: Applies to applications built using common existing web frameworks.
    \item \textbf{Semantic transparency}: Fully answers queries that comply with the policy and blocks queries that do not (rather than providing partial, and potentially misleading, results).
    \item \textbf{Policy expressiveness}: Supports a wide range of policies. 
    \item \textbf{Low overhead}: Has limited impact on page load time.
\end{enumerate}

We implement this approach in \name, a system that enforces a data-access policy at runtime by intercepting SQL queries issued by the application, verifying that they comply with the policy, and blocking those that do not.
We assume non-compliant queries are rare in production (having been mostly eliminated in testing), and focus on efficiently checking compliant queries.
\begin{mychange}
\name{} expects the developer to insert access checks as usual; it merely ensures that the checks are adequate.%
\end{mychange}

A \name{} policy consists of SQL view definitions that specify what information can be accessed by a given user,  although the application still issues queries against the base tables as usual (rather than against the views).  Under this setting, a query is compliant if it \emph{never} reveals---for any underlying dataset---more information than the views do, a well-studied property in databases called \emph{query determinacy}~\cite{Nash10:determinacy}.

While determinacy characterizes the compliance of one query in isolation, it is too restrictive in the context of web applications, which typically issue multiple queries when serving a request.
In this setting, what queries can be allowed often depends on the result of previous queries in the same web request.  Thus, we extend determinacy to take a \emph{trace} of previous queries and their responses, a novel extension we call \emph{trace determinacy}, and use that as the criterion for compliance.

To verify compliance, \name{} frames trace determinacy as an SMT formula and checks it using SMT solvers.
As we later explain, a solver returns an unsatisfiability proof when a query is compliant, and a test demonstrating a violation otherwise.

This basic method, while correct, is impractically slow as it invokes solvers on every query.  Thus, we use a \emph{decision cache} to record compliant queries (with traces) so that future occurrences need not be rechecked.
But caching \emph{exact} queries and traces would be ineffective: a query is usually specific to the user and page visited, and so is unlikely to occur many times.

Thus, to increase cache hit rate, we implement a novel generalization mechanism which, given a compliant query-trace pair, extracts a small set of assumptions on the query and trace that alone would guarantee compliance.  These assumptions
are cached in the form of a \emph{decision template}, which will apply to all future query-trace pairs that meet those assumptions.
\name{} generates decision templates by progressively relaxing a query and trace while maintaining compliance,
with the help of solver-generated unsat cores~\cite[\S~11.8]{barrett18:smt}.
It does not cache noncompliance results, which we expect to be rare in production as they typically indicate application/policy bugs.

We applied \name{} to three existing applications---\diaspora~\cite{diaspora}, Spree~\cite{spree}, and Autolab~\cite{autolab}---and found that it imposes an overhead of
\begin{mychange}
\SIrange{2}{12}{\percent}
\end{mychange}
to the median page load time when compliance decisions are cached.

\name{} has some important limitations. It assumes that the application obtains all of its information through SQL queries visible to \name or from a caching layer or file system mediated by \name.
It also supports only a subset of SQL and is at the mercy of solver performance and unsat-core size.

\ifextended
    \name{} is open source at \url{https://github.com/blockaid-project}.
\fi

%% file: related.tex
The subject of data-access control has been studied by many.
We compare our approach to prior ones along our goals (\Cref{sec:intro}).

\paragraph{Static verification.}
Several systems have been proposed to statically verify that application code can only issue compliant queries;
examples include Swift~\cite{chong07:swift}, \textsc{SELinks}~\cite{corcoran09:selinks}, UrFlow~\cite{chlipala10:urflow}, and \textsc{Storm}~\cite{lehmann21:storm}.
These systems incur no run-time overhead and can be more precise than
\name as they analyze source code.
However, they typically require using a specialized language
or framework like Jif~\cite{myers99:jflow} or Ur/Web~\cite{chlipala10:ur},
sacrificing compatibility with common web frameworks.

\paragraph{Query modification.}
A popular run-time approach is query modification~\cite{stonebraker74:acl}: replacing secret values returned by a query with placeholders (or dropping any rows containing secrets).
This is implemented in commercial databases~\cite{Browder02:vpd,msft21:sql_rls} and academic works like Hippocratic databases~\cite{Agrawal02:hippocratic}, Jacqueline~\cite{yang16:jacqueline},
Qapla~\cite{mehta17:qapla}, and multiverse databases~\cite{marzoev19:multiverse}.
While this approach allows programmers to issue queries without regard to policies,
it lacks semantic transparency as it can alter query semantics in unexpected ways and return misleading results~\cite{Wang07:correctness,rizvi04:rewriting,Guarnieri14:optimal}.%

Furthermore, many of the query modification mechanisms use row- and cell-level policies (e.g., SQL Server RLS and DDM, Oracle VPD).
As we discuss in \Cref{sec:discussion}, this row/cell-level format is less expressive than \name{}'s view-based scheme. 

\paragraph{View-based access control.}
Many databases allow creating views and granting access to views and tables.
Although identical in expressiveness to \name, this mechanism requires queries to explicitly use view names instead of table names (like \texttt{Users}).
\begin{mychange}%
This marks a significant deviation from regular web programming, as programmers must now sort out which views to use for each query.
In contrast, \name{} allows queries to be issued against the base tables directly.%
\end{mychange}

While some prior work has studied view-based compliance of queries issued against base tables~\cite{bender13:fine,bender14:explainable},
they only check single queries, while \name{} checks a query in the context of a trace, a crucial feature for supporting web applications.


%% file: design.tex
\subsection{Application Assumptions and Threat Model}
\begin{figure}
    \centering
    \includegraphics[width=\columnwidth]{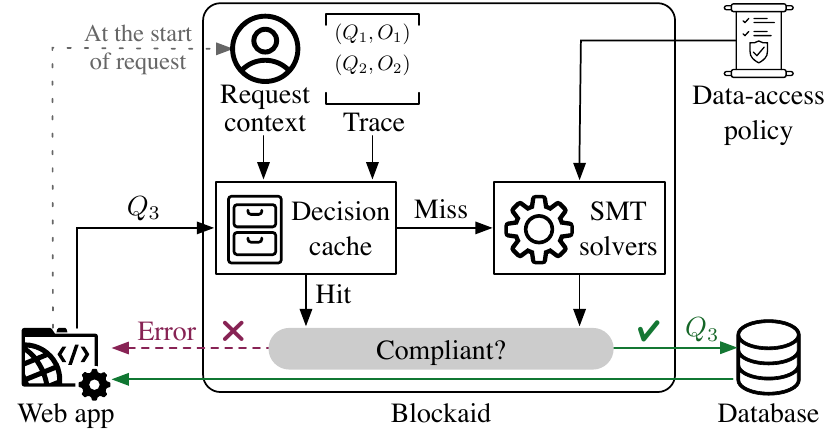}
    \vspace{-.25in}
    \caption{An overview of \name{} (for a single web request).}\label{fig:overview}%
\end{figure}

\name targets web applications that store data in a SQL database.
We assume that a user is logged in and that the current user's identifier is stored in a \emph{request context}.
The application can access the database and the request context when serving a request;
each request is handled independently from others.
We assume that the application authenticates the user correctly,
and that the correct request context is passed to \name~(\Cref{sec:design:arch}).

A \emph{data-access policy} dictates, for a given request context, what information in the database
is \emph{accessible} and what is \emph{inaccessible}.
We treat the database schema and the policy itself as public knowledge
and assume that the user cannot use side channels to circumvent policies.
We enforce policies on database \emph{reads} only, as done in prior work~\cite{agrawal05:privacy,
bender14:explainable,bender13:fine,brodsky00:secure,halder10:fine,
lefevre04:hippocratic,rizvi04:rewriting,shi09:soundness,
stonebraker74:acl,Wang07:correctness,marzoev19:multiverse}.
\begin{mychange}
Ensuring the integrity of updates, while important, is orthogonal to our goal and is left to future work.%
\end{mychange}

\subsection{System Overview}\label{sec:design:arch}
\name{} is a SQL proxy that sits between the application and the database~(\Cref{fig:overview}).
It takes as input
\begin{inparaenum}[(1)]
    \item a database schema (including constraints), and
    \item a data-access policy specified as database views~(\Cref{sec:spec}),
\end{inparaenum}
and checks query compliance for each web request separately.
For each web request, it maintains a \emph{trace} of queries issued so far and their results;
the trace is cleared when the request ends.
\begin{mychange}
\name{} assumes that the results returned by queries in the trace are not altered till the end of the request.
\end{mychange}

When a web request starts, the application sends its request context to \name{}.
Then, every SQL query from the application traverses \name, which attempts to verify that the query is \emph{compliant}---i.e., it can be answered using accessible information only.
To do so, \name{} checks the decision cache for any similar query has been determined compliant previously.
If not, it encodes noncompliance as an SMT formula~(\Cref{sec:checking}) and checks its satisfiability using several SMT solvers in parallel~(\Cref{sec:impl}).

If a query is compliant, \name forwards it to the database unmodified.
In case of a cache miss, \name also extracts and caches a decision template (\Cref{sec:cache}).
Finally, it appends the query and its result to the trace.
If verification fails, \name blocks the query by raising an error to the application.

Although our core design assumes that all sensitive information is stored in the relational database, \name supports limited compliance checking for two other common data sources:
\begin{enumerate}
    \item If the application stores database-derived data in a {\bf caching layer} (e.g., Redis),
        the programmer can annotate a cache key pattern with SQL queries from which the value can be derived.
        \name can then intercept each cache read and verify the compliance of the queries associated with the key.
    \item If the application stores sensitive data in the {\bf file system},
        it can generate hard-to-guess names for these files and store the file names in a database column protected by the policy.
\end{enumerate}

\name's basic requirement is soundness: preventing the revelation of inaccessible information (formalized in \Cref{sec:spec:ni}).
However, it may reject certain behaviors that do not violate the policy (\Cref{sec:discussion}), although such false rejections never arose in our evaluation (\Cref{sec:eval}).
\begin{mychange}%

We end by emphasizing two aspects of \name's operation:
\begin{enumerate}
    \item \name{} has \emph{no visibility into or control over} the application (except by blocking queries).
        So it must assume that \emph{any} data fetched by the application will be shown to the user.
    \item \name{} has \emph{no access} to the database except by observing query results---%
        it cannot issue additional queries of its own.
\end{enumerate}%
\end{mychange}

\subsection{Application Requirements}\label{sec:design:code}
For use with \name, an application must:
\begin{enumerate}
    \item Send the request context to \name at the start of a request and signal \name to clear the trace at the end;
    \item Handle rejected queries cleanly (although a web server's default behavior of returning HTTP~500 often suffices); and,
    \item Not query data that it does not plan on revealing to the user.
\end{enumerate}

\begin{mychange}%
    Existing applications often violate the third requirement.
    For example, when a user views an order on a Spree e-commerce site, the order is fetched from the database, and only then does Spree check, in application code, that the user is allowed to view it.
    To avoid spurious errors from \name,
    such applications must be modified
    to fetch only data known to be accessible.%
\end{mychange}

%% file: specification.tex

Throughout the paper, we will use as a running example a calendar application with the following database schema:
\begin{align*}
    \textit{Users}(&
        \underline{\textit{UId}}, \textit{Name}) \\
    \textit{Events}(&
    \underline{\textit{EId}}, \textit{Title}, \textit{Duration}) \\
    \textit{Attendances}(&
    \underline{\textit{UId}, \textit{EId}},\textit{ConfirmedAt})
\end{align*}
where primary keys are \underline{underlined}.
The request context consists of a parameter $\textit{MyUId}$ denoting
the \textit{UId} of the current user.

\begin{mylisting}
\caption{Example policy view definitions $V_1$ to $V_4$ for the calendar application. \texttt{?MyUId} refers to the current user ID.}\label{listing:views}
    \begin{mdframed}[skipabove=0em]\small%
        \begin{enumerate}[itemsep=1ex]
            \item
                \begin{compactsql}
SELECT * FROM Users
                \end{compactsql}
                \textit{Each user can view the information on all users.}
            \item
                \begin{compactsql}
SELECT * FROM Attendances
WHERE  UId = ?MyUId
                \end{compactsql}
                \textit{Each user can view their own attendance information.}
            \item
                \begin{compactsql}
SELECT * FROM Events
WHERE  EId IN (SELECT EId
               FROM   Attendances
               WHERE  UId = ?MyUId)
                \end{compactsql}
            \textit{Each user can view the information on events they attend.}
            \item
                \begin{compactsql}
SELECT * FROM Attendances
WHERE  EId IN (SELECT EId
               FROM   Attendances
               WHERE  UId = ?MyUId)
                \end{compactsql}
            \textit{Each user can view all attendees of the events they attend.}
        \end{enumerate}
    \end{mdframed}
\end{mylisting}

\subsection{Specifying Policies as Views}\label{sec:spec:spec}
\begin{mychange}%
A policy is a collection of SQL queries that, together, define what information a user is allowed to access.
Each query is called a \emph{view definition} and can refer to parameters from the request context.
As an example, \Cref{listing:views} shows four view definitions, $V_1$--$V_4$;
we denote this policy as $\V=\{V_1,V_2,V_3,V_4\}$.

Notationally, for a view~$V$ and a request context~$\ctx$,
we write $V^{\ctx}$ to denote $V$ with its parameters replaced with values in~$\ctx$.
We often drop the superscript when the context is apparent.%
\end{mychange}

\subsection{Compliance to View-based Policy}\label{sec:spec:compliance}
\begin{mychange}%
Under a policy consisting of view definitions, \name can allow an application query to go through \emph{only if} it is certain
that the query's result is \emph{uniquely determined} by the views.
In other words, an allowable query must be answerable using accessible information alone.
If a query's output \emph{might} depend on information outside the views,
\name must block the query.

\begin{example}
    Let $\textit{MyUId} = 2$.
    The following query selects the names of everyone whom the user attends an event with:
    \begin{lstlisting}
SELECT DISTINCT u.Name
FROM   Users u
       JOIN Attendances a_other
         ON a_other.UId = u.UId
       JOIN Attendances a_me
         ON a_me.EId = a_other.EId
WHERE a_me.UId = 2
    \end{lstlisting}
    Looking at \Cref{listing:views}, this query can always be answered by combining $V_4$,
    which reveals the \textit{UId} of everyone whom the user attends an event with, with $V_1$,
    which supplies the names associated with these \textit{UId}'s.
    Hence, \name allows it through.
\end{example}%
\end{mychange}

\begin{mychange}%
This query above is allowed \emph{unconditionally} because it is answerable using the views on \emph{any} database instance.
More commonly, queries are allowed \emph{conditionally} based on what \name has learned about the current database state,
given the trace of prior queries and results in the same web request.

\begin{example}\label{ex:trace-compliance}
    Again, let $\textit{MyUId} = 2$.  Consider the following sequence of queries issued while handling one web request:
    \smallskip
        \begin{enumerate}
            \item
                \begin{compactsql}
SELECT * FROM Attendances
WHERE UId = 2 AND EId = 5
                \end{compactsql}
                \begin{rrlist}
                    \item \texttt{(UId=2, EId=5, ConfirmedAt="05/04 1pm")}
                \end{rrlist}
        \end{enumerate}
        \begin{enumerate}[start=2]
            \item
                \begin{compactsql}
SELECT Title FROM Events WHERE EId = 5
                \end{compactsql}
        \end{enumerate}
    \smallskip
    The application first queries the user's attendance record for Event~\#5---an unconditionally allowed query---%
    and receives one row, indicating the user is an attendee.
    It then queries the title of said event.
    This is allowed because $V_3$ reveals the information on all events attended by the user.
    More precisely, the trace limits our scope to only databases where the user attends Event~\#5.
    Because the second query is answerable using~$V_3$ \emph{on all such databases},
    it is conditionally allowed given the trace.
\end{example}%
\end{mychange}

\begin{mychange}%
Context is important here: the second query cannot be safely allowed if it were issued in isolation.

\begin{example}\label{ex:noncompliant}
    Suppose instead that the application issues the following query by itself:
    \begin{lstlisting}
SELECT Title FROM Events WHERE EId = 5
    \end{lstlisting}
    \name must block this query because it is not answerable using~$\V$ on a database where the user does not attend Event~\#5.
    Whether or not the user \emph{actually} is an attendee of the event is irrelevant:
    The application, not having queried the user's attendance records,
    cannot be certain that the query is answerable using accessible information alone.
    This differs from alternative security definitions~\cite{Guarnieri14:optimal,Koutris12:pricing,Zhang05:authorization-views}
    where a policy enforcer can allow a query after inspecting additional information in the database that has not been fetched by the application.
\end{example}%
\end{mychange}

\begin{mychange}%
\begin{definition}
    A \emph{trace}~$\T$ is a sequence $(Q_1,O_1),\ldots,(Q_n,O_n)$ where each $Q_i$ is a query and each $O_i$ is a collection of tuples. 
\end{definition}%

Such a trace denotes that the application has issued queries $Q_1,\ldots,Q_n$ and received results $O_1,\ldots,O_n$ from the database.

We now motivate the formal definition of query compliance given a trace (using colors to show correspondence between text and equations).
Consider any two databases that are:
\begin{itemize}
    \item \textcolor{tol1}{Equivalent in terms of accessible data} (i.e., they differ only in information outside the views), and
    \item \textcolor{tol2}{Consistent with the observed trace} (i.e., we consider only databases that \emph{could} be the one the application is querying).
\end{itemize}
\name{} must ensure that such two databases are indistinguishable to the user---%
by allowing only queries that \textcolor{tol8}{produce the same result on both databases}.
\end{mychange}


\begin{definition}\label{def:compliance}
    \begin{mychange}%
    Let $\ctx$ be a request context, $\V$ be a set of views, and $\T=\{(Q_i,O_i)\}_{i=1}^{n}$ be a trace.
    A query~$Q$ is \emph{$\ctx$-compliant} to~$\V$ given~$\T$ if for every pair of databases $D_1,D_2$
that conform to the database schema and constraints,\footnote{We will henceforth use ``schema'' to mean both schema and constraints, and rely on the database and/or the web framework to enforce the constraints.}
    and satisfy:
    \begin{align}
        \textcolor{tol1}{V^{\ctx}(D_1)} &\; \textcolor{tol1}{=V^{\ctx}(D_2)}, & \textcolor{tol1}{(\forall V\in\V)} \label{eqn:compliance1} \\
        \textcolor{tol2}{Q_i(D_1)} &\;\textcolor{tol2}{= O_i}, & \textcolor{tol2}{(\forall 1\leq i\leq n)} \label{eqn:compliance2} \\
        \textcolor{tol2}{Q_i(D_2)} &\;\textcolor{tol2}{= O_i}, & \textcolor{tol2}{(\forall 1\leq i\leq n)} \label{eqn:compliance3}
    \end{align}
    we have $\textcolor{tol8}{Q(D_1)=Q(D_2)}$.
    We will simply say \emph{compliant} if the context is clear.%
    \end{mychange}
\end{definition}



We call \Cref{def:compliance} \emph{trace determinacy} because it extends the classic notion of query determinacy~\cite{Nash10:determinacy,Segoufin05:determinacy} with the trace.
Query determinacy is undecidable even for conjunctive views and queries~\cite{Gogacz15,Gogacz16};
trace determinacy must also be undecidable in the same scenario.
Although several decidable cases have been discovered for query determinacy~\cite{Nash10:determinacy,Afrati11,Pasaila11}, they are not expressive enough for our use case.
A promising direction is to identify classes of views and queries that capture common web use cases and for which trace determinacy is decidable.

\subsection{From Query Compliance to Noninterference}\label{sec:spec:ni}
\begin{mychange}%
\name's end goal is to ensure that an application's output depends only on information accessible to the user.
In relation to this goal, query compliance (\Cref{def:compliance}) satisfies two properties, making it the right criterion for \name{} to enforce:
\begin{enumerate}
    \item \textbf{Sufficiency}:
        As long as \emph{only compliant queries} from the application are let through,
        there is no way for an execution's outcome to be influenced by inaccessible information.
    \item \textbf{Necessity}:
        Any enforcement system that makes per-query decisions based solely on the query and its preceding trace
        \emph{cannot safely allow any non-compliant query} without the risk of the application revealing inaccessible information.
\end{enumerate}
Before stating and proving these properties formally, let us first model our target applications, enforcement systems, and goals.

We model a web request handler as a program $\Prog(\ctx,\req,D)$
that maps a request context~$\ctx$, an HTTP request~$\req$, and a database~$D$ to an HTTP response.%
\footnote{For simplicity, we assume $\Prog$ is a pure function---deterministic, terminating, and side-effect free---%
although this assumption can be relaxed through standard means from information-flow control~\cite[\S~2]{HedinS12:ifc}.}
A program that abides by a policy~$\V$ satisfies a \emph{noninterference} property~\cite{Cohen77,Goguen82}
stating that its output depends only on the inputs that the user has access to---%
namely, $\ctx$, $\req$, and $V^{\ctx}(D)$ for each $V\in\V$.
The formal definition follows from a similar intuition as \Cref{def:compliance}.

\begin{definition}
    A program~$\Prog$ satisfies \emph{noninterference} under policy~$\V$ if the following condition holds:
    \begin{align*}
        \NI_{\V}(\Prog) &\coloneqq \forall \ctx, \req, D_1, D_2\ldotp \\
                               & \qquad\left[ \forall V\in\V\ldotp V^{\ctx}(D_1)=V^{\ctx}(D_2) \right] \\
        &\qquad\qquad \implies \Prog(\ctx,\req,D_1) = \Prog(\ctx,\req,D_2).
    \end{align*}
\end{definition}

An enforcement system must ensure that any program running under it satisfies noninterference.
We now model such a system that operates under \name's assumptions.

\begin{definition}
    An \emph{enforcement predicate} is a mapping from a request context, a query, and a trace to an allow/block decision:
    \[
        \EnforcePred(\ctx, Q, \T) \to \{\cmark, \xmark\}.
    \]
\end{definition}

\begin{definition}
    Let $\Prog(\ctx,\req,D)$ be a program and $\EnforcePred$ be an enforcement predicate.
    We define the program~$\Prog$ \emph{under enforcement using~$\EnforcePred$} as a new program $\Prog^{\EnforcePred}(\ctx,\req,D)$ that simulates every step taken by the original program~$\Prog$,
    except that it maintains a trace~$\mathcal{T}$ and blocks any query~$Q$ issued by~$\Prog$ where $\EnforcePred(\ctx,Q,\mathcal{T})=\xmark$ by immediately returning an error.
\end{definition}

Note that $\Prog^{\EnforcePred}$ evaluates~$\EnforcePred$ only on traces in which every query has been previously allowed by~$\EnforcePred$ given its trace prefix.

\begin{definition}
    Given a request context~$\ctx$,
    we say that a trace $\T=\{(Q_i, O_i)\}_{i=1}^{n}$ is \emph{prefix $\EnforcePred$-allowed} if for all $1\leq i\leq n$,
    \[
        \EnforcePred(\ctx,Q_i,\T[1..i-1])=\cmark.
    \]
\end{definition}


\begin{definition}
    A predicate~$\EnforcePred$ \emph{correctly enforces} policy~$\V$ if:
    \[
        \forall\Prog\ldotp~\NI_{\V}(\Prog^{\EnforcePred}).
    \]
\end{definition}

We are ready to state the sufficiency-and-necessity theorem, whose proof is left to \refAppendix{sec:proofs}.
Like before, we use colors to link a statement to its explanation.
\begin{theorem}\label{thm:compliance}
    Let $\V$ be a set of views and $\EnforcePred$ be a predicate.
    \begin{enumerate}
        \item Suppose $\EnforcePred(\ctx,Q,\T)=\cmark$ \textcolor{tol1}{only when $Q$ is $\ctx$-compliant} to~$\V$ given~$\T$.
            Then $\EnforcePred$ \textcolor{tol2}{correctly enforces $\V$}.
        \item Suppose $\EnforcePred$ \textcolor{tol7}{correctly enforces $\V$}.
            Then \textcolor{tol8}{for any} request context~$\ctx$, query~$Q$, and prefix $\EnforcePred$-allowed trace~$\T$ \textcolor{tol8}{such that $\EnforcePred(\ctx,Q,\T)=\cmark$,
            $Q$ is $\ctx$-compliant} to~$\V$ given~$\T$.
    \end{enumerate}
\end{theorem}

To unpack, \Cref{thm:compliance} says:
\begin{inparaenum}[(1)]
\item as long as an enforcement predicate \textcolor{tol1}{ensures query compliance}, it \textcolor{tol2}{correctly enforces the policy} on applications (i.e., sufficiency); and
\item for a predicate to \textcolor{tol7}{correctly enforce the policy}, it must \textcolor{tol8}{ensure query compliance} (i.e., necessity).
\end{inparaenum}
Thus, query compliance can be regarded as the ``projection'' of application noninterference onto \name's lens, making it the ideal criterion to enforce.%
\end{mychange}

%% file: checking.tex
Having defined view-based policy and compliance, we now introduce how \name{} verifies compliance using SMT solvers.


\subsection{Translating Noncompliance to SMT}\label{sec:checking:smt}
\name{} verifies query compliance by framing \emph{noncompliance} (i.e, the negation of \Cref{def:compliance}) as an SMT formula and checking its satisfiability---%
a query is compliant if and only if the formula is \emph{unsatisfiable}.
We use a straightforward translation based on Codd's theorem~\cite{Codd72:relational-completeness},
which states, informally,
that relational algebra under set semantics is equivalent in expressiveness to first-order logic (FOL).
Relational algebra has five operators---projection, selection, cross product, union, and difference---%
and tables are interpreted as \emph{sets} of rows (i.e., no duplicates).
Under this equivalence, tables are translated to predicates in FOL,
and operators are implemented using existential quantifiers, conjunctions, disjunctions, and negations.

\begin{example}\label{ex:query-fol}
    Let us translate into FOL the following query~$Q$ executed on a database~$D$:
\begin{lstlisting}
SELECT e.EId, e.Title
FROM   Events e, Attendances a
WHERE  e.EId = a.EId AND a.UId = 2
\end{lstlisting}
    Let $E^D(\cdot,\cdot,\cdot)$ and $A^D(\cdot,\cdot,\cdot)$ be FOL predicates representing
    the $\textit{Events}$ and $\textit{Attendances}$ table in the database~$D$ in:
    \begin{align*}
        \sfq^D(\sfx_e,\sfx_t) &\coloneqq \exists x_d, x_u, x_e', x_c\ldotp
        E^D(\sfx_e, \sfx_t, x_d) \land A^D(x_u,x_e',x_c) \\
            & \qquad \land \sfx_e=x_e' \land x_u = 2.
    \end{align*}
    $\sfq^D(\sfx_e,\sfx_t)$ encodes the statement $(\sfx_e,\sfx_t)\in Q(D)$,
    i.e., that the row~$(\sfx_e,\sfx_t)$ is returned by $Q$ on database~$D$.
    Note that $\sfq^D$~is not a logical symbol, but merely a shorthand for the right-hand side.
\end{example}

\begin{example}
    We now present the noncompliance formula for a single query~$Q$ with respect to~$\V$ from \Cref{sec:spec:spec}. 
    Let $\sfv_1^{D_i},\ldots,\sfv_4^{D_i}$ and~$\sfq^{D_i}$ encode the views and query on database~$D_i$ ($i=1,2$) in FOL.
    The desired formula would then be the conjunction of:
    \begin{align*}
        \forall \bfx\ldotp \sfv_1^{D_1}(\bfx) &\leftrightarrow \sfv_1^{D_2}(\bfx), & (V_1(D_1)=V_1(D_2)) \\
                            &\vdots \\
        \forall \bfx\ldotp \sfv_4^{D_1}(\bfx) &\leftrightarrow \sfv_4^{D_2}(\bfx), & (V_4(D_1)=V_4(D_2)) \\
        \exists \bfx\ldotp \sfq^{D_1}(\bfx) &\not\leftrightarrow \sfq^{D_2}(\bfx), & (Q(D_1)\neq Q(D_2))
    \end{align*}
    where $\bfx$ denotes a sequence of fresh variables.
    Database constraints and consistency with a trace can be encoded similarly.
\end{example}


\subsection{Handling Practical SQL Queries}\label{sec:checking:sql}
The encoding of relational algebra into logic, while straightforward, fails to cover real-world SQL due to two semantic gaps:
\begin{enumerate}
    \item While the encoding assumes that relational algebra is evaluated under \emph{set semantics},
        in practice databases use a mix of set, bag, and other semantics when evaluating queries.\footnote{
        For example, a SQL \texttt{SELECT} clause can return duplicate rows,
        but the \texttt{UNION} operator removes duplicates.
        }
    \item SQL operations like aggregation and sorting have no corresponding operators in relational algebra.
\end{enumerate}

For \name to bridge these gaps, it must first assume that database tables contain no duplicate rows.
This is generally the case for web applications as object-relational mapping libraries like Active~Record~\cite{ActiveRecord} and Django~\cite{Django} add a primary key for every table.
Given this assumption, \name rewrites complex SQL into \emph{basic queries} that map directly to relational algebra.

\subsubsection{Basic SQL Queries}\label{sec:checking:sql:basic}
\begin{definition}\label{def:basic}
    A \emph{basic} query is either a \texttt{SELECT}-\texttt{FROM}-\texttt{WHERE} query that never returns duplicate rows,
    or a \texttt{UNION} of \texttt{SELECT}-\texttt{FROM}-\texttt{WHERE}~clauses
    (the \texttt{UNION} always removes duplicates).\footnote{The \texttt{MINUS}~operator is not used in our applications and is omitted.}
\end{definition}
A basic query on duplicate-free tables maps to relational algebra under set semantics, and so can be directly translated to FOL.
To ensure a \texttt{SELECT} query is basic, we check it against these sufficient conditions for returning no duplicate rows:
\begin{itemize}
    \item It contains the \texttt{DISTINCT} keyword or ends in \texttt{LIMIT 1}; or
    \item It projects unique key column(s) from every table in~\texttt{FROM}, e.g.,
        \mintinline{sql}{SELECT UId, Name FROM Users}; or
    \item It is constrained by uniqueness in its \texttt{WHERE}~clause---e.g.:
\begin{lstlisting}
SELECT e.EId
FROM   Events e, Attendances a
WHERE  e.EId = a.EId AND a.UId = 2
\end{lstlisting}
For this query to return multiple copies of $x$,
the database must contain multiple rows of the form $\textit{Attendances}(2,x,\texttt{?})$;
this is ruled out by the uniqueness constraint on $(\textit{UId},\textit{EId})$.
\end{itemize}


\begin{mychange}%
In our experience, policy views can typically be written as basic queries directly---%
e.g., for \Cref{listing:views} we can frame $V_3$ and $V_4$ as equivalent basic queries
by replacing subqueries with joins and using the inner join transformation from \Cref{sec:checking:sql:rewrite}.
\end{mychange}

\subsubsection{Rewriting Into Basic Queries}\label{sec:checking:sql:rewrite}
\begin{mychange}%
When the application issues a query~$Q$, \name{} attempts to rewrite it into a basic query~$Q'$ and verify its compliance instead.
Ideally, $Q'$ would be equivalent to $Q$, but when this is not possible,
\name{} produces an \emph{approximate} $Q'$ that reveals \emph{at least as much} information as $Q$~does.%
\footnote{It suffices to guarantee that $Q$ can be computed from the result of~$Q'$.}
Such approximation preserves soundness but may sacrifice completeness,
although it caused no false rejections in our evaluation.
We now explain how to rewrite several types of queries encountered in practice.%
\end{mychange}

\paragraph{Inner joins.}
A query of the form:
\begin{lstlisting}
SELECT ... FROM R1
INNER JOIN R2 ON C1 WHERE C2
\end{lstlisting}
is equivalently rewritten as the basic query:
\begin{lstlisting}
SELECT ... FROM R1, R2 WHERE C1 AND C2
\end{lstlisting}

\paragraph{Left joins on a foreign key.}
Consider a query of the form:
\begin{lstlisting}
SELECT ... FROM R1
LEFT JOIN R2 ON R1.A = R2.B WHERE ...
\end{lstlisting}
\begin{mychange}%
If \texttt{R1.A} is a foreign key into \texttt{R2.B},
then every row in \texttt{R1} matches at least one row in \texttt{R2}.
In this case, the left join can be equivalently written as an inner join, which is handled as above.%
\end{mychange}

\paragraph{Order-by and limit.}
\name{} adds any \texttt{ORDER BY} column as an output column
and then discards the \texttt{ORDER BY} clause.
It also discards any \texttt{LIMIT} clause but, when adding the query to the trace,
uses a modified condition $O_i\subseteq D(Q_i)$ (instead of ``$=$'') to indicate that
it may have observed a partial result.

\paragraph{Aggregations.}
\begin{mychange}%
\name{} turns
\mintinline{sql}{SELECT SUM(A) FROM R}
into
\mintinline{sql}{SELECT PK, A FROM R},
where \texttt{PK} is table \texttt{R}'s primary key.
By projecting the primary key in addition to \texttt{A},
the rewritten query reveals the multiplicity of the values in \texttt{A}---necessary for computing \lstinline{SUM(A)}---%
without returning duplicate rows.%
\end{mychange}

\paragraph{Left joins that project one table.}
Left joins of the form:
\begin{lstlisting}
SELECT DISTINCT A.* FROM A
LEFT JOIN B ON C1 WHERE C2
\end{lstlisting}
can be equivalently rewritten to the basic query:
\begin{lstlisting}
(SELECT A.* FROM A
 INNER JOIN B ON C1 WHERE C2)
UNION
(SELECT * FROM A WHERE C3)
\end{lstlisting}
where \texttt{C3} is obtained by replacing each occurrence of \texttt{B.?}
with \texttt{NULL} in \texttt{C2} and simplifying the resulting predicate.%
\footnote{As long as \texttt{C2}~contains no negations, it is safe to treat a
\texttt{NULL} literal as \texttt{FALSE} when propagating through or short-circuiting
\texttt{AND} and \texttt{OR} operators.}
\begin{mychange}
The first subquery covers the rows in~\texttt{A} with at least one match in \texttt{B},
and the second subquery covers those with no matches.%
\end{mychange}

\paragraph{Feature not supported.}
The SQL features not supported include \texttt{GROUP BY},
\texttt{ANY}, \texttt{EXISTS}, etc.,
although they can also be formulated / approximated using basic queries.
In the future we plan to leverage other formalisms~\cite{cheung13:qbs,chu17:hottsql,veanes09:query_exp,veanes10:qex,wang17:synth,chu18:axiom,wang18:equiv} to model complex SQL
semantics more precisely.

\subsection{Optimizations and SMT Encoding}\label{sec:checking:opt}
We end this section with several optimizations for compliance checking and some notes on the SMT encoding.

\paragraph{Strong compliance.}
\begin{mychange}
We define a stronger notion of compliance, which we found SMT solvers can verify more efficiently.

\begin{definition}\label{def:strong-compliance}
    A query~$Q$ is \emph{strongly $ctx$-compliant} to policy~$\V$ given trace $\{(Q_i,O_i)\}_{i=1}^{n}$
    if for each pair of databases $D_1,D_2$ that conform to the schema and satisfy:
    \begin{align}
        V^{\ctx}(D_1) &\subseteq  V^{\ctx}(D_2), & (\forall V\in\V) \label{eqn:strongCompliance1} \\
        Q_i(D_1) &\supseteq O_i, & (\forall 1\leq i\leq n) \label{eqn:strongCompliance2}
    \end{align}
    we have $Q(D_1)\subseteq Q(D_2)$.
\end{definition}


\begin{theorem}\label{thm:strong}
    If $Q$~is strongly compliant to~$\V$ given trace $\T$, then $Q$ is also compliant to~$\V$ given~$\T$.
\end{theorem}
\begin{proof}
    Let $Q$ be strongly compliant to~$\V$ given~$\T$.
    To show that $Q$ is also compliant, let $D_1, D_2$ be databases that satisfy \Crefrange{eqn:compliance1}{eqn:compliance3} from the compliance definition.
    These imply the strong compliance assumptions (\Cref{eqn:strongCompliance1,eqn:strongCompliance2}),
    and so we have $Q(D_1)\subseteq Q(D_2)$.
    By symmetry, we also have $Q(D_2)\subseteq Q(D_1)$.
    Putting the two together, we conclude $Q(D_1)=Q(D_2)$, showing $Q$ to be compliant to~$\V$ given~$\T$.
\end{proof}

For faster checking, \name{} verifies strong compliance rather than compliance; by \Cref{thm:strong}, soundness is preserved.
However, there are scenarios where a query is compliant but \emph{not} strongly compliant (see \refAppendix{sec:strong});
such queries will be falsely rejected by \name.
This did not pose a problem in practice as we found the two notions to coincide for every query encountered in our evaluation.
\end{mychange}

\paragraph{Fast accept.}
Given a view \sql{SELECT C1, ...,} \sql{Ck FROM R}, any query that references only
columns \texttt{R.C1}, \dots, \texttt{R.Ck} must be compliant and is accepted without SMT solving.

\paragraph{Trace pruning.}
Queries that returns many rows can inflate the trace and slow down the solvers.
Fortunately, often times only few of the rows matter to a later query's compliance.
We thus adopt a trace-pruning heuristic: when checking a query $Q$, look for any previous query has returned over ten rows,
and keep only those rows that contain the first occurrence of a primary-key value (e.g., user ID) appearing in~$Q$.
This heuristic is sound, but may need to be adapted for any application where our premise for pruning does not hold.

\paragraph{SQL types and predicates.}
To model SQL types, we use SMT's uninterpreted sorts, which we found to yield better performance than
theories of integers, strings, etc.
We support logical operators \texttt{AND} and \texttt{OR},
comparison operators \texttt{<}, \texttt{<=}, \texttt{>}, \texttt{>=},
and operators \texttt{IN}, \texttt{NOT IN},%
\footnote{We only support \texttt{IN} and \texttt{NOT IN} with a list of values, not with a subquery.}
\texttt{IS NULL}, and \texttt{IS NOT NULL}.
We model \texttt{<} as an uninterpreted relation with a transitivity axiom.

\paragraph{NULLs.}
We model \texttt{NULL} using a two-valued semantics of SQL~\cite[\S~6]{guagliardo17:sql}
by (1)~designating a constant in each sort as \texttt{NULL},
and (2)~taking \texttt{NULL} into account when implementing SQL operators.
\begin{mychange}
For example, the SQL predicate \texttt{x=y} translates into the following SMT formula:
$x=y \land x\neq \textit{null} \land y\neq\textit{null}$.
\end{mychange}

%% file: caching.tex
While SMT solvers can check a wide range of queries, doing so often takes 100s of milliseconds per query.
As a page load can depend on tens of queries, this overhead can add up to \emph{seconds}.

\begin{mylisting*}
    \caption{An example query with trace from the calendar application and a decision template generated from it.}
    \vspace{-1.5ex}
    \begin{submylisting}[b]{.48\textwidth}
        \caption{
            Example query with trace ($\textit{UId} = 1$).
        }\label{fig:calendar-trace}%
        \begin{mdframed}[skipabove=0em]\small%
        \begin{enumerate}
            \item
                \begin{minted}{sql}
SELECT * FROM Users WHERE UId = 1
                \end{minted}
                \begin{rrlist}
                    \item \texttt{(UId=1, Name="John Doe")}
                \end{rrlist}
            \item
                \begin{minted}{sql}
SELECT * FROM Attendances
WHERE UId = 1 AND EId = 42
                \end{minted}
                \begin{rrlist}
                    \item \texttt{(UId=1, EId=42, ConfirmedAt="05/04 1pm")}
                \end{rrlist}
        \end{enumerate}
        \rule{\textwidth}{0.4pt}
        \begin{enumerate}[start=3]
            \item
                \begin{minted}{sql}
SELECT * FROM Events WHERE EId = 42
                \end{minted}
        \end{enumerate}
        \end{mdframed}
    \end{submylisting}\hfill%
    \begin{submylisting}[b]{.48\textwidth}
        \caption{The decision template generated by \name.}\label{fig:template-ex}%
        \begin{mdframed}[skipabove=0em]\small
            \begin{enumerate}
                \item
                \begin{minted}{sql}
SELECT * FROM Attendances
WHERE UId = |\prm{?MyUId}| AND EId = |\prm{?0}|
                \end{minted}
                \begin{rrlist}
                    \item \verb!(UId = !\prm{\texttt{?MyUId}}\verb!, EId = !\prm{\texttt{?0}}\verb!, ConfirmedAt = !\prm{\texttt{*}}\verb!)!
                \end{rrlist}
            \end{enumerate}
            \rule{\textwidth}{0.4pt}
            \begin{qlist}
                \item
                \begin{minted}{sql}
SELECT * FROM Events WHERE EId = |\prm{?0}|
                \end{minted}
            \end{qlist}
        \end{mdframed}
    \end{submylisting}
\end{mylisting*}

To alleviate this overhead, \name{} aims to reduce solver calls by caching compliance decisions.
Naively, once query~$Q$ is deemed compliant given trace~$\T$,
we could record $(Q,\T)$ and allow future occurrences without re-invoking the solvers.

However, this proposal is unlikely to be effective because the number of distinct $(Q,\T)$~pairs can be unbounded.
For example, an application can issue as many queries of the form \sql{SELECT * FROM Users WHERE UId = ?} as there are users in the system.
Therefore, requiring an exact query-trace match for a cache hit would result in a low cache hit rate.

Fortunately, while an application can issue an unbounded number of distinct queries,
it only exhibits a finite number of truly different behaviors.
For example, the query sequences generated by requests for two different calendar events
are likely identical in structure while differing only in parameters (e.g., event ID).
If one sequence is compliant, we can \emph{generalize} this knowledge to conclude that the other is also compliant.

This generalization problem is the central challenge we tackle in this section:
Given a query's compliance with respect to a trace,
how to abstract this knowledge into a \emph{decision template} such that
\begin{inparaenum}[(1)]
\item any query (and its trace) that matches this template is compliant, and
\item the template is general enough to produce matches on similar requests.
\end{inparaenum}
Such a template, once cached, will apply to an entire class of traces and queries.

Decision templates are designed to cache compliant queries only.
Our techniques do not extend to non-compliant queries, which are expected to be rare in production
as they typically indicate bugs in the application or the policy.

Let us start with an example of a decision template.

\subsection{Example}\label{sec:cache:ex}
Suppose a user with $\textit{UId}=1$ requests Event \#42 in the calendar application,
resulting in the application issuing a sequence of SQL queries.
Consider the third query, shown in~\Cref{fig:calendar-trace}.
As we explained in~\Cref{ex:trace-compliance}, Query~\#3 is compliant because
Query~\#2 has established that the user attends the event.

\name{} aims to abstract this query (with trace) into a decision template that applies to another user viewing a different event.
\Cref{fig:template-ex} shows such a template; the notation says:
If each query-output pair above the line has a match in a trace~$\mathcal{T}$,
then any query of the form below the line is compliant given~$\mathcal{T}$.
This particular template states: after it is determined that
user~$x$ attends event~$y$, user~$x$ can view event~$y$
for any $x$ and $y$.

Compared with the concrete query and trace, this template
\begin{inparaenum}[(1)]
\item omits Query~\#1, which is immaterial to the compliance decision; and
\item replaces the concrete values with parameters.
\end{inparaenum}
Occurrences of~\prm{\texttt{?0}} here constrain the event ID fetched by the query to equal the previously checked event ID.
We use~\prm{\texttt{*}} to denote a fresh parameter, i.e., any arbitrary value is allowed.

We now dive into how \name{} extracts such a decision template from a concrete query and trace.
But before we do so, let us first define what a decision template is, what it means for a template to have a match, and what makes a ``good'' template.

\subsection{Definitions and Goals}
For convenience, from now on we will denote a trace as a set of query-\emph{tuple} pairs $\{(Q_i,t_i)\}_{i=1}^{n}$,
where each $t_i$ is \emph{one} of the rows returned by~$Q_i$.
A query that returns multiple rows is represented as multiple such pairs.
\begin{mychange}%
This change of notation is permissible because under strong compliance (\Cref{def:strong-compliance}),
we no longer take into account the \emph{absence} of a returned row.

\begin{definition}
    We say a trace~$\T=\{(Q_i,t_i)\}_{i=1}^{n}$ is \emph{feasible} if there exists a database~$D$ such that $t_i\in Q_i(D)$ for all $1\leq i\leq n$.
\end{definition}

\begin{definition}
    A \emph{decision template} $\D[\bfx,\bfc]$, where $\bfc$ denotes variables from the request context and $\bfx$ a sequence of variables disjoint from $\bfc$,
    is a triple $(Q_{\D},\T_{\D},\Phi_{\D})$ where:
    \begin{itemize}
        \item $Q_{\D}$ is the \emph{parameterized query}, whose definition can refer to variables from~$\bfx \cup \bfc$;
        \item $\T_{\D}$ is the \emph{parameterized trace}, whose queries and tuples can refer to variables from~$\bfx\cup\bfc$; and
        \item $\Phi_{\D}$, the \emph{condition}, is a predicate over $\bfx\cup\bfc$.
    \end{itemize}
    We will often denote a template simply by $\D$ if the variables are either unimportant or clear from the context.
\end{definition}

As we later explain, $\Phi_{\D}$ represents any extra constraints that a template imposes on its variables (e.g., $\prm{?0} < \prm{?1}$).

\begin{definition}
    A \emph{valuation}~$\nu$ over a collection of variables~$\bfy$ is a mapping from $\bfy$ to constants (including \texttt{NULL}),
    extended to objects that contain variables in~$\bfy$.
    For example, given a parameterized query~$Q$, $\nu(Q)$ denotes $Q$ with each occurrence of variable $y\in\bfy$ substituted with $\nu(y)$.
\end{definition}

\begin{definition}
    Let $\D[\bfx,\bfc]=(Q_{\D},\T_{\D},\Phi_{\D})$ be a decision template, $\ctx$ be a request context,
    $\T$ be a trace, and $Q$ be a query.
    We say that $\D$ \emph{matches $(Q,\T)$ under $\ctx$} if there exists a valuation~$\nu$ over $\bfx\cup\bfc$ such that:
    \begin{itemize}
        \item $\nu(\bfc)=\ctx$,
        \item $\nu(Q_{\D})=Q$,
        \item $\left(\nu(Q_j), \nu(t_j)\right)\in \T$ for all $\left(Q_j,t_j\right)\in\T_{\D}$, and
        \item $\nu(\Phi_{\D})$ holds.
    \end{itemize}
\end{definition}
\end{mychange}

\begin{mychange}%
\begin{example}
    \Cref{fig:template-ex} can be seen as a stylized rendition of a decision template $\D[\bfx,\bfc]$
    where $\bfx=(x_0,x_1)$---$x_0$ denoting \prm{\texttt{?0}} and $x_1$ denoting the occurrence of \prm{\texttt{*}}---and $\bfc=(\textit{MyUId})$;
    $Q_{\D}$ and $\T_{\D}$ are as shown below and above the line;
    and $\Phi_{\D}$ is the constant $\top$, meaning the template imposes no additional constraints on the variables.\footnote{Technically, this template requires $\textit{MyUId}\neq\texttt{NULL}\land x_0\neq \texttt{NULL}$.  We omitted this condition in \Cref{fig:template-ex} because we assume the user ID parameter and the \textit{Attendances} table's \textit{EId} column are both non-NULL.}
    Under the request context $\textit{MyUId}=1$, this template matches the query and trace in \Cref{fig:calendar-trace}
    via the valuation $\{x_0\mapsto 42, x_1\mapsto \texttt{"05/04 1pm"}, \textit{MyUId}\mapsto 1\}$.
\end{example}

We are interested only in templates that imply compliance.

\begin{definition}
    A decision template $\D$ is \emph{sound} with respect to a policy~$\V$ if for every request context~$\ctx$,
    whenever $\D$ matches $(Q,\T)$ under $\ctx$, $Q$ is strongly $ctx$-compliant to~$\V$ given~$\T$.
\end{definition}

\name{} can verify that a template is sound via the following theorem derived from strong compliance (\Cref{def:strong-compliance}):

\begin{theorem}
    A decision template $\D[\bfx,\bfc]=(Q_{\D},\T_{\D},\Phi_{\D})$ is sound with respect to a policy~$\V$ if and only if:
     \begin{align*}
         &\forall \bfx, \bfc, D_1, D_2\ldotp \\
         &\quad\left.
         \begin{aligned}
             \Phi_{\D} \\
             \forall V\in\V\ldotp V(D_1) \subseteq  V(D_2) \\
             \forall (Q_i,t_i)\in\T_{\D}\ldotp t_i \in Q_i(D_1)
         \end{aligned}
         \right\} \implies Q_{\D}(D_1)\subseteq Q_{\D}(D_2).
     \end{align*}
\end{theorem}

For a compliant query~$Q$ (with trace~$\T$) that misses the cache, there often exist many sound templates that match $(Q,\T)$.
But all such templates are not equal---we prefer the more \emph{general} ones, those that match a wider range of \emph{other} queries and traces.

\begin{definition}
    A template~$\D_1$ is \emph{at least as general as} a template $D_2$ if for every query~$Q$ and feasible trace~$\T$,
    if $\D_2$ matches $(Q,\T)$, $\D_1$ also matches $(Q,\T)$.
\end{definition}

Thus, \name{} aims to generate a decision template that
\begin{inparaenum}[(1)]
    \item is sound,
    \item matches $(Q,\T)$, and
    \item is general enough for practical purposes.
\end{inparaenum}
We now explain how this is achieved.
\end{mychange}

\subsection{Generating Decision Templates}\label{sec:cache:core}
\begin{mychange}%
\name{} starts from the trivial template $D_0=(Q,\T,\top)$, which is sound but not general,
and generalizes it in two steps:
\begin{enumerate}
    \item Minimize the trace~$\T$ to retain only those $(Q_i,t_i)$~pairs that are required for $Q$'s compliance (\Cref{sec:cache:core:min}).
    \item Replace each constant in the trace and query with a fresh variable,
        and then generate a weak condition~$\Phi$ over the variables that guarantees compliance (\Cref{sec:cache:core:cond}).
\end{enumerate}%
\end{mychange}

\subsubsection{Step One: Trace Minimization}\label{sec:cache:core:min}
\begin{mychange}%
\name{} begins by finding a minimal sub-trace of~$\T$ that preserves compliance.
It removes each $(Q_i,t_i)\in\T$ and, if $Q$ is no longer compliant, adds the element back.
For example, for \Cref{fig:calendar-trace} this step removes Query~\#1.
Denote the resulting minimal trace by $\Tmin$ and let decision template $\D_1=(Q,\Tmin,\top)$.

\begin{proposition}
    $\D_1$ is sound, matches $(Q,\T)$, and is at least as general as $\D_0$.
\end{proposition}

As an optimization, \name{} starts the minimization
from the sub-trace that the solver has actually used to prove compliance.
It extracts this information from a solver-generated \emph{unsat core}~\cite[\S~11.8]{barrett18:smt}---%
a subset of clauses in the formula that remains unsatisfiable even with all other clauses removed.
If we attach \emph{labels} to the clauses we care about,
a solver will identify all labels in the unsat core when it proves the formula unsatisfiable.

To get an unsat core, \name{} uses the following formula:
\begin{align*}
                    && V^{\ctx}(D_1) &\subseteq  V^{\ctx}(D_2), & (\forall V\in\V) \\
    [\textit{LQ}_i] && t_i &\in Q_i(D_1), & (\forall (Q_i,t_i)\in\T) \\
                    && Q(D_1) &\not\subseteq Q(D_2),
\end{align*}
where the clause asserting the $i$\textsuperscript{th} trace entry is labeled $\textit{LQ}_i$.
If $Q$ is compliant, the solver returns as the unsat core a set~$S$ of labels.
\name{} ignores any $(Q_i,t_i)\in\T$ for which $\textit{LQ}_i\not\in S$.%
\end{mychange}

\subsubsection{Interlude: Model Finding for Satisfiable Formulas}
A common operation in template generation is to remove parts of a formula and re-check satisfiability.
A complication arises when the formula turns satisfiable---%
while solvers are adept at proving unsatisfiability, they often fail on satisfiable formulas.%
\footnote{For example, finite model finders in CVC4~\cite{reynolds13:finite} and Vampire~\cite{reger16:finite} often time out or run out of memory on tables with only tens of columns.}

\begin{mychange}%
To solve these formulas faster,
we observe that
they are typically satisfied by databases with small tables.
We thus construct SMT formulas to directly seek such ``small models'' by representing each table not as an uninterpreted relation,
but as a conditional table~\cite{Imielinski84} whose size is bounded by a small constant.

A conditional table generalizes a regular table by
\begin{inparaenum}[(1)]
\item allowing variables in its entries, and
\item associating with each row with a \emph{condition}, i.e., a Boolean predicate for whether the row exists.
\end{inparaenum}
For example, a \textit{Users} table with a bound of~2 appears as:
\begin{center}
\begin{tabular}{cccc}
    \toprule
    \textit{UId} & \textit{Name} & Exists? \\
    \midrule
    $x_{u,1}$ & $x_{n,1}$ & $b_1$ \\
    $x_{u,2}$ & $x_{n,2}$ & $b_2$ \\
    \bottomrule
\end{tabular}
\end{center}
where each entry and condition is a fresh variable, signifying that the table is not constrained in any way other than its size.%

Queries on condition tables are evaluated via an extension of the relational algebra operators~\cite[\S~7]{Imielinski84}.
This allows queries to be encoded into SMT without using quantifiers or using relation symbols for tables.%
\footnote{To avoid using quantifiers in these formulas, we drop the transitivity axiom for the uninterpreted less-than relation (\Cref{sec:checking:opt}).}
For example, the query
\lstinline{SELECT Name FROM Users WHERE UId = 5} can be written as:
\end{mychange}
\[
    \sfq(\sfx_n) \coloneqq
    \bigvee_{i=1}^{2}
    \left(
        x_{u,i}=5 \land x_{n,i}=\sfx_n \land b_i
    \right).
\]
We found that such formulas could be solved quickly by Z3.

After \name{} generates an unsat core as described in \Cref{sec:cache:core:min},
it switches to using bounded formulas (i.e., ones that use conditional tables instead of uninterpreted relations) for the remainder of template generation.
\name{} sets a table's bound to one plus the number of rows required to produce the sub-trace induced by the unsat core;%
\footnote{If the bounds are too small for a database to produce the trace, the resulting formula will be unsatisfiable regardless of compliance.}
it relies on the solvers to produce small unsat cores to keep formula sizes manageable.

Care must be taken because using bounded formulas breaks soundness---%
a query compliant on small tables might not be on larger ones.
Therefore, after a decision template is produced \name{} verifies its soundness on the unbounded formula, and if this fails, increments the table bounds and retry.

\subsubsection{Step Two: Find Value Constraints}\label{sec:cache:core:cond}
Taking the template $\D_1=(Q,\Tmin,\top)$ from Step~1,
\name{} generalizes it further by abstracting away the constants.
To do so, \name{} \emph{parameterizes} $\Tmin$ and $Q$ by replacing each occurrence of a constant with a fresh variable.
We use a superscript ``p'' to denote the parameterized version of a query, tuple, or trace.
\Cref{fig:parameterized} shows~$\paramd{\Tmin}$ and~$\paramd{Q}$ from our example.
As an optimization, \name{} assigns the same variable (e.g., \texttt{x0}) to locations that are guaranteed by SQL semantics to be equal.

\begin{mylisting}
    \caption{Parameterization and candidate atoms for~\Cref{fig:calendar-trace}.}\label{fig:base-template}%
    \begin{submylisting}{\columnwidth}
        \vspace{-1.5ex}
        \caption{Parameterized trace~$\paramd{\Tmin}$ and query~$\paramd{q}$.}\label{fig:parameterized}%
        \begin{mdframed}[skipabove=0em]\small
            \begin{enumerate}[leftmargin=*,start=2]
                \item
                \begin{minted}{sql}
SELECT * FROM Attendances
WHERE UId = |\prm{x0}| AND EId = |\prm{x1}|
                \end{minted}
                \begin{rrlist}
                    \item \verb!(UId = !\prm{\texttt{x0}}\verb!, EId = !\prm{\texttt{x1}}\verb!, ConfirmedAt = !\prm{\texttt{x2}}\verb!)!
                \end{rrlist}
            \end{enumerate}
            \rule{\textwidth}{0.4pt}
            \begin{enumerate}[leftmargin=*,start=3]
                \item
                    \begin{minted}{sql}
SELECT * FROM Events WHERE EId = |\prm{x3}|
                    \end{minted}
            \end{enumerate}
        \end{mdframed}%
    \end{submylisting}
    \begin{submylisting}{\columnwidth}
        \vspace{1.5ex}
        \caption{Candidate atoms (with symmetric duplicates removed).}\label{fig:constraints}%
        \begin{mdframed}[skipabove=0em]\small
            \begin{minipage}[t]{.36\columnwidth}
            Form \texttt{x = v}:
            \begin{itemize}
                \item \texttt{MyUId = 1}
                \item \texttt{x0 = 1}
                \item \texttt{x1 = 42}
                \item \texttt{x2 = "05/04 1pm"}
                \item \texttt{x3 = 42}
            \end{itemize}
            \end{minipage}\hfill%
            \begin{minipage}[t]{.29\columnwidth}
            Form \texttt{x = x'}:
            \begin{itemize}
                \item \texttt{MyUId = x0}
                \item \texttt{x1 = x3}
            \end{itemize}
            \end{minipage}\hfill%
            \begin{minipage}[t]{.29\columnwidth}
            Form \texttt{x < x'}:
            \begin{itemize}
                \item \texttt{MyUId < x1}
                \item \texttt{MyUId < x3}
                \item \texttt{x0 < x1}
                \item \texttt{x0 < x3}
            \end{itemize}
            \end{minipage}
        \end{mdframed}%
    \end{submylisting}
\end{mylisting}

\begin{mychange}%
\name{} must now generate a condition $\Phi$ such that the resulting template
$\D_2=(\paramd{Q}, \paramd{\Tmin}, \Phi)$ meets our goals.
It picks as $\Phi$ a conjunction of atoms from a set of \emph{candidate atoms}.
Let $\bfx$ denote all variables generated from parameterization, and let $\nu$ map~$\bfx$ to the replaced constants
and $\bfc$ to the current context~$\ctx$.

\begin{definition}\label{def:candidate}
    The set of \emph{candidate atoms} is defined as:
    \[
        C = \bigcup
            \begin{cases}
                \{ \texttt{x = v} &\mid x\in\bfx\cup\bfc, v=\nu(x)\neq\texttt{NULL} \} \\
                \{ \texttt{x IS NULL} &\mid x\in\bfx\cup\bfc, \nu(x)=\texttt{NULL} \} \\
                \{ \texttt{x = x'} &\mid x,x'\in\bfx\cup\bfc, \nu(x)=\nu(x')\neq\texttt{NULL} \} \\
                \{ \texttt{x < x'} &\mid x,x'\in\bfx\cup\bfc, \nu(x)<\nu(x') \}
            \end{cases}.
    \]
    (We write atoms in \texttt{monospace font} to distinguish them from mathematical expressions.
    Following SQL, the ``\texttt{=}'' in an atom implies that both sides are non-\texttt{NULL}.)
\end{definition}

Note that all candidate atoms hold on $Q$ and $\Tmin$.
\name{} now selects a subset that not only guarantees compliance, but also imposes relatively few restrictions on the variables.

\begin{definition}
    With respect to $\paramd{Q}$ and $\paramd{\Tmin}$, a subset of atoms $C_0\subseteq C$ is \emph{sound} if the decision template $(\paramd{Q}, \paramd{\Tmin}, \bigwedge C_0)$ is sound.
    ($\bigwedge C_0$ denotes the conjunction of atoms in~$C_0$.)
\end{definition}

\begin{definition}
    Let $C_1,C_2\subseteq C$.
    We say that $C_2$ is \emph{at least as weak as} $C_1$ (denoted $C_1\preceq C_2$) if $\bigwedge C_1 \implies \bigwedge C_2$,
    and that $C_2$ is \emph{weaker} than $C_1$ if $C_1\preceq C_2$ but $C_2\not\preceq C_1$.
\end{definition}

\begin{example}
    \Cref{fig:constraints} shows all the candidate atoms from \Cref{fig:parameterized} (after omitting symmetric ones in the \texttt{x = x'} group).
    Consider the following two subsets of atoms:
    \begin{align*}
        C_1 &=
        \begin{Bmatrix}
            \texttt{MyUId = x0}, & \texttt{x1 = 42}, & \texttt{x3 = 42}
        \end{Bmatrix},\\
        C_2 &=
        \begin{Bmatrix}
            \texttt{MyUId = x0}, & \texttt{x1 = x3}
        \end{Bmatrix}.
    \end{align*}
    While both are sound, $C_2$ is preferred over $C_1$ as it is weaker and thus applies in more scenarios.
    In fact, $C_2$ is \emph{maximally} weak: there exists no subset that is both sound and weaker than~$C_2$.
\end{example}

Ideally, \name{} would produce a maximally weak sound subset of~$C$ to use as the template condition, but finding one can be expensive.
It thus settles for finding a subset that is weak enough for practical generalization. It does so in three steps.%
\end{mychange}

\textbf{First}, as a starting point, \name{} generates a minimal unsat core of the formula:
\begin{align*}
                    && V^{\ctx}(D_1) &\subseteq  V^{\ctx}(D_2), & (\forall V\in\V) \\
                    && \paramd{t_i} &\in \paramd{Q_i}(D_1), & (\forall (\paramd{t_i}, \paramd{Q_i})\in \paramd{\Tmin}) \\
    [\textit{LC}_i] && c_i &, & (\forall c_i\in C) \\
                    && \paramd{Q}(D_1) &\not\subseteq \paramd{Q}(D_2).
\end{align*}
Let $\Ccore$ denote the atoms whose label appears in the unsat core.
For example, $\Ccore=\Set{\texttt{MyUId = x0}, \texttt{x1 = 42}, \texttt{x3 = 42} }$.

\textbf{Second}, it augments $\Ccore$ with other atoms that are implied by it:
$\Caug=\Set{c\in C | \bigwedge \Ccore \implies c}$.
In our example,
\begin{align*}
    \Caug &= \Ccore\cup \Set{\texttt{x1 = x3}} \\
                   &=
                   \begin{Bmatrix}
                       \texttt{MyUId = x0}, & \texttt{x1 = 42}, & \texttt{x3 = 42}, & \texttt{x1 = x3}
                   \end{Bmatrix}.
\end{align*}
$\Caug$ enjoys a closure property: if $C_0\subseteq \Caug$ and $C_0\preceq C_1$, then $C_1\subseteq \Caug$.
In particular, $\Caug$ contains a maximally weak sound subset of~$C$.
Thus, \name{} focuses its search within $\Caug$.




\textbf{Finally}, as a proxy for weakness,
\name{} finds a \emph{smallest} sound subset of $\Caug$, denoted $\Csmall$, breaking ties arbitrarily.
It does so using the \marco{} algorithm~\cite{liffiton13:marco,previti13:emus,liffiton16:marco} for minimal unsatisfiable subset enumeration,
modified to enumerate from small to large and to stop after finding the first sound subset.
In our example, the algorithm returns $\Csmall=\Set{\texttt{MyUId=x0}, \texttt{x1=x3}}$ of cardinality two, which is also a maximally weak subset (even though this might not be the case in general).%
\footnote{For example, $\Set{\texttt{x < y}, \texttt{x < z}}$ is strictly weaker than $\Set{\texttt{x < y}, \texttt{y < z}}$ even though the two sets have the same cardinality.}
Nevertheless, searching for a smallest sound subset has produced templates that generalize well in practice.

\medskip
\begin{mychange}%
At the end, \name{} produces the decision template:
\[
    \D_2[\bfx,\bfc] = \left(\paramd{Q}, \paramd{\Tmin}, \bigwedge \Csmall\right).
\]
\begin{proposition}
    $\D_2$ is sound, matches $(Q,\T)$, and is at least as general as $\D_1$.
\end{proposition}
As an optimization, whenever $\bigwedge \Csmall \implies x=y$ for $x,y\in\bfx\cup\bfc$, \name{} replaces $x$ with $y$ in the template.
This is how, e.g., in \Cref{fig:template-ex} \prm{?0} appears in both the trace and the query.%
\end{mychange}


\subsubsection{Optimizations}\label{sec:cache:core:opt}
We implement two optimizations that improve the performance of template generation
and the generality of templates.

\paragraph{Omit irrelevant tables.}  Given trace~$\T$ and query~$Q$,
we call a table \emph{relevant} if (1)~it appears in~$\T$ or~$Q$,
or (2)~the table appears on the right-hand side of a database constraint of the form~$Q_1\subseteq Q_2$,
given that a relevant table appears on the left.%
\footnote{Every constraint encountered in our evaluation can be written in the form
$Q_1\subseteq Q_2$, including primary-key, foreign-key, and integrity constraints.}
\name{} sets the size bounds of irrelevant tables to zero, reducing formula size while preserving compliance.

\paragraph{Split \texttt{IN}.}
A query~$Q$ that contains ``$c~\texttt{IN}~(x_1,x_2,\ldots,x_n)$''
often produces a template with a long trace.
If $Q$ is a basic query that does not contain the \texttt{NOT}~operator,
it can be split into $q_1,\ldots,q_n$ where $q_i$ denotes~$Q$ with the \texttt{IN}-construct
substituted with~$c = x_i$, such that $Q \equiv q_1 \cup \ldots \cup q_n$.
If $q_1,\ldots,q_n$ are all compliant then so is~$Q$,
and so \name{} checks the subqueries instead.
This is usually fast because $q_2,\ldots,q_n$ typically match the decision template generated from~$q_1$.
If any~$q_i$ is not compliant, \name{} reverts to checking~$Q$ as a whole.

This optimization also improves generalization.
Suppose $Q'$ has structure identical to~$Q$ but a different number of \texttt{IN}~operands.
It would not match a template generated from~$Q$,
but its split subqueries~$q_i'$ could match the template from~$q_1$.

\subsection{Decision Cache and Template Matching}\label{sec:cache:cache}
\begin{mychange}%
\name{} stores decision templates in its \emph{decision cache}, indexing them by their parameterized query using a hash map.
When checking a query~$Q$, \name{} lists all templates whose parameterized query matches~$Q$; for each such template, it uses recursive backtracking (with pruning optimizations) to search for a valuation that results in a match.
This simple method proves efficient in practice as the templates tend to be small.%
\end{mychange}

%% file: impl.tex
We implemented \name{} as a Java Database Connectivity~(JDBC) driver that interposes on an underlying connection.
It thus supports only applications on the JVM and runs within the web server, although our design allows it to reside elsewhere (e.g., in the database).
The JDBC driver accepts custom commands that
\begin{inparaenum}[(1)]
    \item set the request context,
    \item clear the context and the trace, and
    \item check an application cache read.
\end{inparaenum}

\name{} parses SQL using Apache Calcite~\cite{begoli18:calcite} and caches parser outputs.
To check compliance,
it uses Z3's Java binding~\cite{deMoura12:z3-java} to generate formulas in SMT-LIB~2 format~\cite{barrett17:smtlib}
and invokes an ensemble of solvers in parallel.
Our ensemble consists of Z3~\cite{demoura08:z3} (v4.8.12) and
\cvc~\cite{barbosa22:cvc5} (v0.3) using default configurations,
and Vampire~\cite{kovacs13:vampire} (v4.6.1) using six configurations from its CASC portfolio.%
\footnote{\url{https://github.com/vprover/vampire/blob/master/CASC/Schedules.cpp}.}
The ensemble is killed as soon as any solver finishes.
If a query is not compliant, or all solvers time out after~\SI{5}{\second},
\name{} throws a Java \texttt{SQLException}.

To generate decision templates, \name{} uses the same ensemble to produce the initial unsat core (\Cref{sec:cache:core:min}),
but kills the ensemble only when a solver returns a small core of up to 3~labels (subject to timeout).
It uses only Z3 on bounded formulas.

Our prototype does not verify that queries return no duplicate rows and does not look at any \texttt{ORDER BY} columns.
We manually ensured that queries in our evaluation return no duplicates and do not reveal inaccessible information through~\texttt{ORDER BY}.

%% file: eval.tex
We use \name to enforce data-access policies on three existing open-source web applications written in Ruby on Rails:
\begin{itemize}
    \item \textbf{\diaspora}~\cite{diaspora}: a social network with 850~k users.
    \item \textbf{Spree}~\cite{spree}: an e-commerce app used by 50+ businesses.
    \item \textbf{Autolab}~\cite{autolab}: a course management app used at 20~schools.
\end{itemize}
For each application, we devised a data-access policy,
modified its code to work with \name,
and measured its performance.

In summary: \name{} imposes overheads of \SI{2}{\percent}--\SI{12}{\percent} to median page load time when compliance decisions are cached;
the decision templates produced by \name{} generalize to other entities (users, etc.); and no query was falsely rejected in our benchmark.
Instructions for reproducing our experiments can be found in \Cref{sec:artifact}.

\begin{table}\small
    \caption{Summary of schemas, policies, and code changes.}\label{tbl:policies}\centering
    \vspace{-1ex}
    \begin{tabular}{lrrr}
        \toprule
             & \textbf{\diaspora} & \textbf{Spree} & \textbf{Autolab} \\
             \midrule
        \multicolumn{4}{l}{\textbf{Schema \& Policy}} \\
        \quad\# Tables modeled & 35 / 52 & 46 / 93 & 17 / 28 \\
        \quad\# Constraints & 108 & 122 & 51 \\
        \quad\# Policy views & 108 & 84 & 57 \\
        \quad\# Cache key patterns & 0 & 11 & 3 \\
        \multicolumn{4}{l}{\textbf{Code Changes (LoC)}} \\
        \quad Boilerplate & 12 & 17 & 12 \\
        \quad Fetch less data & 6 & 26 & 38 \\
        \quad SQL feature & 1 & 3 & 5 \\
        \quad Parameterize queries & 0 & 18 & 32 \\
        \quad File system checking & 0 & 0 & 9 \\
        \quad \textit{Total} & 19 & 64 & 96 \\
        \bottomrule
    \end{tabular}
\end{table}

\begin{table*}
\caption{Application benchmark.
    For a page we list the \underline{page URL} followed by other URLs fetched (URLs for assets are excluded).
    When compliance decisions are cached, \name incurs up to \SI{12}{\percent} overhead to the median PLT over the modified applications.}\label{tbl:benchmark}
\small\centering
\input{figures/plt.tex}
\end{table*}

\subsection{Constraints, Policies, and Annotations}\label{sec:eval:app}
\Cref{tbl:policies} summarizes the constraints and policies for database tables queried in our benchmark,
including any necessary application-level constraints (e.g., a reshared post is always public in \diaspora).
Spree and Autolab use the Rails cache, and we annotate their cache key patterns with queries (\Cref{sec:design:arch}).

\begin{mychange}%
Once a policy is given, transcribing it into views was straightforward.
The more arduous task lied in divining the intended policy for an application, by studying its source code and interacting with it on sample data.
This effort was complicated by edge cases in policies---e.g., a Spree
item at an inactive location is inaccessible \emph{except} when filtering for backorderable variants.
Such edge cases had to be covered using additional views.

To give a sense of the porting effort, writing the Spree policy took one of us roughly a month.
However, this process would be easier for the developer of a new application, who has a good sense of what policies are suitable and can create policies while building the application, amortizing the effort over time.%
\end{mychange}

When writing the Autolab policy,
we uncovered two access-check bugs in the application:
\begin{inparaenum}[(1)]
    \item a persistent announcement (one shown on all pages of a course) is displayed regardless of whether it is active on the current date, and
    \item an unreleased handout is hidden on its course page but can be downloaded from its assignment page.
\end{inparaenum}
This experience corroborates the difficulty of making every access check airtight,
especially for code bases that enjoy fewer maintenance resources.

\subsection{Code Modifications}\label{sec:eval:change}
Our changes to application code fall into five categories:
\begin{enumerate}
    \item \textbf{Boilerplate}: We add code that sends the request context to \name at request start and clears the trace at request end.
    \item \textbf{Fetch less data}: We modify code to not fetch potentially sensitive data unless it will be revealed to the user;
        some of these changes use the \texttt{lazy\_column} gem~\cite{lazycolumn}.
    \item \textbf{SQL features}: We modify some queries to avoid SQL features not supported by \name (e.g., general left joins) without altering application behavior.
    \item \textbf{Parameterize queries}: We make some queries parameterized so that \name{} can effectively cache their parsing results.
        Most changes are mechanical rewrites of queries with comparisons,
        as idiomatic ways of writing comparisons~\cite{railsLt}
        cause query parameters to be filled within Rails.
    \item \textbf{File system checking}: Autolab uses files to store submissions; the file name are always accessible but the content is inaccessible during an exam.  We modify it to store the submission content under a randomly generated file name and restrict access to the file name in the database (\Cref{sec:design:arch}).
\end{enumerate}

The code changes are summarized also in \Cref{tbl:policies},
which omits configuration changes, adaptations for JRuby, and experiment code.
The changes range from 19 to 96 lines of code.

\subsection{Experiment Setup and Benchmark}\label{sec:eval:setup}
We deploy each application on an Amazon EC2 c4.8xlarge instance running Ubuntu~18.04.
Because our prototype only supports JVM applications~(\Cref{sec:impl}),
we run the applications using JRuby~\cite{jruby} (v9.3.0.0),
a Ruby implementation atop the JVM (we use OpenJDK~17).
In Rails's database configuration, we turn on \texttt{prepared\_statements}
so that Rails issues parameterized queries in the common case.%
\footnote{\begin{mychange}%
In case a Rails query is not fully parameterized (e.g., due to the use of raw SQL), it gets parameterized by \name as described in \Cref{sec:cache:core:cond}.%
\end{mychange}}
The applications run atop the Puma web server
over HTTPS behind NGINX (which serves static files directly),
and stores data in MySQL (and, if applicable, Redis) on the same instance.
To reduce variability, all measurements are taken from a client on the same instance.

For each application, we picked five page loads that exercise various behaviors (\Cref{tbl:benchmark}).
Each page load can fetch multiple URLs, some common among many pages (e.g., D9, which is the notifications URL).
All queries issued are compliant, and all experiments are performed with the Rails cache populated.

\subsection{Page Load Times}\label{sec:eval:plt}
We start by measuring page load times~(PLTs) using a headless Chrome browser (v96) driven by Selenium~\cite{selenium}.
PLTs are reported as the time elapsed between \texttt{navigationStart}
and \texttt{loadEventEnd} as defined by the \texttt{PerformanceTiming}
interface~\cite{navtime}.
The one exception is the Autolab ``Submission'' page, a file download,
for which we report Chrome's download time instead.
Since the client is on the same VM as the server, 
these experiments reflect the best-case PLT, as clients
outside the instance / cloud are likely to experience higher network latency.

We report PLTs under four settings:
\emph{original} (unmodified application), \emph{modified} (modified {\`a} la~\Cref{sec:eval:change}),
\emph{cached} (modified application under \name{} with every query hitting the decision cache),
and \emph{no cache} (decision caching disabled).
For the first three, we perform 3000~warmup loads before measuring
the PLT of another 3000~loads.
For \emph{no cache}, where each run takes longer, we use 100~warmup loads and 100~measurement loads.

\Cref{tbl:benchmark} shows that when compliance decisions are cached, \name incurs
up to \SI{12}{\percent} overhead to median PLT over the modified application
(and up to \SI{17}{\percent} overhead to P95).
With caching disabled, \name incurs up to $236\times$ higher median PLT.
Compared with the original applications,
the modified versions result in up to \SI{6}{\percent} overhead to median PLT for all pages but
Autolab's ``Submissions'', which suffers a \SI{19}{\percent} overhead.
(The P95 overhead is up to \SI{7}{\percent} for all but two pages with up to \SI{26}{\percent} overhead.)
We will comment on these overheads in the next subsection, where we break down the pages into URLs.

\input{figures/fetch_data.tex}

\begin{figure*}\centering
    \include{figures/fetch}
    \ifextended
        \vspace{-.3in}
    \fi
    \caption{URL fetch latency (median).  With all compliance decisions cached, \name incurs up to \SI[round-mode=places,round-precision=0]{\fetchCachedOverheadMaxPerc}{\percent} overhead over ``modified''.}\label{fig:fetch}
\end{figure*}

\subsection{Fetch Latency}\label{sec:eval:breakdown}


To better understand page load performance,
we separate out the individual URLs fetched by each page (\Cref{tbl:benchmark}), omitting URLs for assets, and measure the latency of fetching each URL (not including rendering time).
The median latencies are shown in \Cref{fig:fetch}.
In addition to the four settings from \Cref{sec:eval:plt}, it includes performance under a ``cold cache'',
where the decision cache is enabled but cleared at the start of each load (100 warmup runs followed by 100 measurements).
When all compliance decisions are cached, \name incurs up to
\SIIntPercent{\fetchCachedOverheadMaxPerc} of overhead
(median \SIIntPercent{\fetchCachedOverheadMedianPerc})
over ``modified''.
In contrast, it incurs
$\num[round-mode=places,round-precision=0]{\fetchColdOverheadMin}\times$--$\num[round-mode=places,round-precision=0]{\fetchColdOverheadMax}\times$
overhead on a cold decision cache,
and
$\num[round-mode=places,round-precision=0]{\fetchNoCacheOverheadMin}\times$--$\num[round-mode=places,round-precision=0]{\fetchNoCacheOverheadMax}\times$
overhead if the decision cache is disabled altogether.

For most URLs, ``cold cache'' is slower than ``no cache'' due to the extra template-generation step.
Two exceptions are D4 and A6, where many structurally identical queries are issued, and so
the performance gain from cache hits \emph{within each URL} offsets the performance hit from template generation.

Compared to the original, the modified \diaspora{} and Spree are up to
\SIIntPercent{\fetchDiasporaSpreeModifiedOverheadMaxPerc} slower (median \SIIntPercent{\fetchDiasporaSpreeModifiedOverheadMedianPerc}),
but Autolab is up to \SIIntPercent{\fetchAutolabModifiedOverheadMaxPerc} slower (median \SIIntPercent{\fetchAutolabModifiedOverheadMedianPerc}).
Autolab routinely reveals partial data on objects that are not fully accessible.
For example, a user can distinguish among the cases where
\begin{inparaenum}[(1)]
    \item a course doesn't exist,
    \item a course exists but the user is not enrolled, and
    \item the user is enrolled but the course is disabled.
\end{inparaenum}
The original Autolab fetches the course in one SQL query
but we had to split it into multiple---checking whether the course exists,
whether it is disabled, etc.---and return an error immediately if one of these checks fails.

In one instance~(S2), the modified version is \SIIntPercent{\fetchSpreeTwoModifiedFasterPerc} faster than the original because we were able to remove queries for potentially inaccessible data that is never used in rendering the URL.

\begin{mylisting*}
    \caption{Two (abridged) decision templates generated for the same parameterized query from Spree.
    \texttt{Token} is a Spree request context parameter identifying the current (possibly guest) user, and \texttt{NOW} is a built-in parameter storing the current time.}\label{fig:bad-template}
    \vspace{-1.5ex}
    \begin{submylisting}[b]{.48\textwidth}
    \caption{This template doesn't fully generalize.}\label{fig:bad1}
    \begin{mdframed}[align=center,skipabove=0em]\small
        \begin{minted}{sql}
SELECT * FROM products WHERE id IN (*, *, *)
        \end{minted}
        \begin{rrlist}
        \item \verb!(id = !\prm{\texttt{?1}}\verb!, available_on < !\prm{\texttt{?NOW}}\verb!, !\\
            \verb! discontinue_on IS NULL, deleted_at IS NULL, *)!
        \end{rrlist}
        \begin{minted}{sql}
SELECT * FROM variants WHERE id IN (*, *, *)
        \end{minted}
        \begin{rrlist}
        \item \verb!(id = !\prm{\texttt{?2}}\verb!, deleted_at IS NULL, !\\
            \verb! discontinue_on IS NULL, product_id = !\prm{\texttt{?1}}\verb!, *)!
        \end{rrlist}
        \rule{\textwidth}{0.4pt}
        \begin{minted}{sql}
SELECT a.* FROM assets a
JOIN variants mv ON a.viewable_id = mv.id
JOIN variants ov ON mv.product_id = ov.product_id
WHERE mv.is_master AND mv.deleted_at IS NULL
  AND a.viewable_type = 'Variant' AND ov.id = |\prm{?2}|
        \end{minted}
    \end{mdframed}
    \end{submylisting}\hfill%
    \begin{submylisting}[b]{.48\textwidth}
    \caption{This template does fully generalize.}\label{fig:bad-fixed}
    \begin{mdframed}[align=center,skipabove=0em]\small
        \begin{minted}{sql}
SELECT * FROM orders WHERE ...
        \end{minted}
        \begin{rrlist}
        \item \verb!(id = !\prm{\texttt{?0}}\verb!, token = !\prm{\texttt{?Token}}\verb!, *)!
        \end{rrlist}
        \begin{minted}{sql}
SELECT * FROM line_items WHERE order_id = |\prm{\texttt{?0}}|
        \end{minted}
        \begin{rrlist}
        \item \verb!(variant_id = !\prm{\texttt{?1}}\verb!, *)!
        \end{rrlist}
        \rule{\textwidth}{0.4pt}
        \begin{minted}{sql}
SELECT a.* FROM assets a
JOIN variants mv ON a.viewable_id = mv.id
JOIN variants ov ON mv.product_id = ov.product_id
WHERE mv.is_master AND mv.deleted_at IS NULL
  AND a.viewable_type = 'Variant' AND ov.id = |\prm{?1}|
        \end{minted}
    \end{mdframed}
    \end{submylisting}
\end{mylisting*}

\subsection{Solver Comparison}\label{sec:eval:solvers}
\begin{figure}
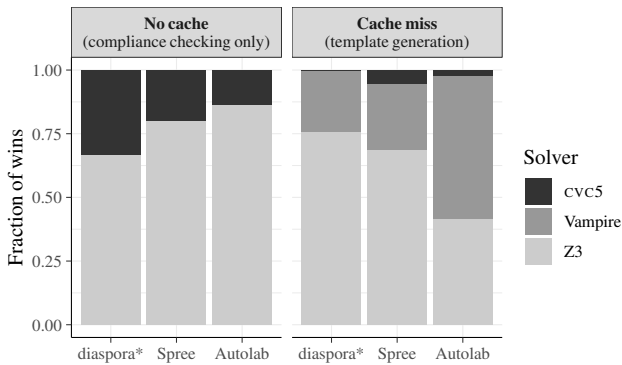

    \centering
    \include{figures/winners}
    \ifextended
        \vspace{-.3in}
    \fi
    \caption{Fraction of wins by each solver. ``Vampire'' covers a portfolio of six configurations (\Cref{sec:impl}).}\label{fig:solvers}
\end{figure}
When a query arrives, \name invokes an ensemble of solvers to check compliance when decision caching is disabled,
and to generate a decision template on a cache miss when caching is enabled.
The \emph{winner}, in the no-cache case, is the first solver to return a decision;
and in the cache-miss case, the first to return a small enough unsat core (\Cref{sec:impl}),
assuming the query is compliant.

\Cref{fig:solvers} shows, in the fetch latency experiments (\Cref{sec:eval:breakdown}), the fraction of wins by each solver in the two cases.
In the no-cache case, the wins are dominated by Z3 followed by \cvc, with none for Vampire.
In the cache-miss case, however, Vampire wins a significant portion of the time.
This is because Z3 and \cvc{} often finish quickly but with large unsat cores,
causing \name to wait till Vampire produces a smaller core.

\subsection{Template Generalization}\label{sec:eval:gen}
We found that the generated decision templates typically generalize to similar requests.
The rest generalize in more restricted scenarios, and none is tied to a particular user ID, post ID, etc.

To illustrate how \name{} might produce a template that fails to generalize fully,
consider a query from Spree's cache key annotations (\Cref{fig:bad-template}).
This query fetches assets for product variants in the user's order.
(Here, the asset of a variant belongs to its product's ``master variant''.)
\Cref{fig:bad1} shows a template that fails to generalize fully, for three reasons.

\textbf{First}, due to the queries with the \texttt{IN} operator in its premise (above the horizontal line),
this template applies only when an order has exactly three variants.
The \texttt{IN}-splitting optimization from \Cref{sec:cache:core:opt} only applies to the query being checked,
and we plan to handle such queries in the premise in future work.

\textbf{Second}, this template constrains the variant to be ``not discontinued'',
which is defined as \mintinline{sql}{discontinue_on IS NULL} or
\mintinline{sql}{discontinue_on >= NOW}.
But because disjunctions are not supported in decision templates,
\name picked only the condition that matches the current variant (\texttt{IS NULL}).

\textbf{Third}, in this example there are multiple justifications for this query's compliance,
and \name happened to pick one that does not always hold in a similar request.
The policy states that a variant's asset can be viewed if it is not discontinued,
or if it is part of the user's order.%
\footnote{This is to allow users to view past purchases that are since discontinued.}
This particular variant in the user's order happens to not be discontinued,
and the template captures the former justification for viewing the asset.
However, it does not apply to variants in the order that \emph{are} discontinued;
indeed, for such variants, \name produces the template in~\Cref{fig:bad-fixed},
which generalizes fully.
We could address this issue by finding multiple decision templates for every query.

Incidentally, inspecting decision templates has helped us expose overly permissive policies.
When writing the Autolab policy, we missed a join condition in a view,
a mistake that became apparent when \name{} generated a template stating that an instructor
for one course can view assignments for \emph{all} courses.
Although manual inspection of templates is not required for using \name{}, doing so can help debug overly broad policies, whose undesirable consequences are often exposed by the general decision templates produced by \name{}.

%% file: figures/plt.tex
\begin{tabular}{lllrrrr}
\toprule
& & & \multicolumn{4}{c}{\textbf{Page Load Time} (median / P95; default unit: ms)} \\
\cmidrule(lr){4-7}
& \textbf{URLs} & \textbf{Description} & Original & Modified & Cached & No cache \\ \midrule
\multicolumn{3}{l}{\textbf{\diaspora}}\\
    \quad{}Simple post &        \underline{D1}, D2, D9 & View a simple post shared with the user.          & 169 / 173 & 169 / 175 & 174 / 179 & \SI{2.5}{s} / \SI{2.6}{s} \\
    \quad{}Complex post &       \underline{D3}, D4, D9 & View a public post with 30~votes and comments.    & 171 / 178 & 171 / 178 & 176 / 183 & \SI{2.6}{s} / \SI{2.7}{s} \\
    \quad{}Prohibited post &    \underline{D5} & Attempt to view an unauthorized post.                     & 32 / 34 & 32 / 34 & 33 / 35 & 262 / 285 \\
    \quad{}Conversation &       \underline{D6}, D9 & View a conversation (5 messages).                     & 253 / 258 & 255 / 262 & 260 / 267 & \SI{2.1}{s} / \SI{2.2}{s} \\
    \quad{}Profile &            \underline{D7}, D8, D9 & View someone's profile (basic info and 3~posts).  & 142 / 148 & 145 / 152 & 150 / 156 & \SI{1.3}{s} / \SI{1.4}{s} \\
\multicolumn{3}{l}{\textbf{Spree}}\\
    \quad{}Account &            \underline{S1}, S6--S8 & View the user's account information.              & 74 / 80 & 76 / 83 & 78 / 84 & 588 / 611 \\
    \quad{}Available item &     \underline{S2}, S6--S8 & View a product for sale.                          & 122 / 133 & 115 / 167 & 122 / 173 & \SI{4.4}{s} / \SI{4.4}{s} \\
    \quad{}Unavailable item &   \underline{S3} & Attempt to view a product no longer for sale.             & 20 / 22 & 21 / 23 & 22 / 24 & 350 / 371 \\
    \quad{}Cart &               \underline{S4}, S6--S8 & View the current shopping cart (3 items).         & 116 / 131 & 118 / 132 & 124 / 137 & \SI{7.6}{s} / \SI{7.7}{s} \\
    \quad{}Order &              \underline{S5}, S6--S8 & View a summary and status of a previous order.    & 160 / 170 & 164 / 174 & 173 / 182 & \SI{39}{s} / \SI{39}{s} \\
\multicolumn{3}{l}{\textbf{Autolab}}\\
    \quad{}Homepage &           \underline{A1} & View a summary of 3 courses enrolled.                     & 56 / 61 & 59 / 64 & 65 / 70 & \SI{1.4}{s} / \SI{1.6}{s} \\
    \quad{}Course &             \underline{A2}, A3 & View summary of one course (15~assignments).          & 84 / 96 & 87 / 101 & 97 / 116 & \SI{3.9}{s} / \SI{4.1}{s} \\
    \quad{}Assignment &         \underline{A4} & View a quiz (incl. 3~submissions and grades).             & 97 / 110 & 103 / 118 & 115 / 138 & \SI{3.5}{s} / \SI{3.6}{s} \\
    \quad{}Submission &         \underline{A5} & Download a previous homework submission.                  & 22 / 26 & 26 / 31 & 27 / 33 & \SI{1.1}{s} / \SI{1.2}{s} \\
    \quad{}Gradesheet &         \underline{A6} & Instructor views grades for 51~enrollees.                 & 456 / 474 & 474 / 493 & 504 / 530 & \SI{72}{s} / \SI{73}{s} \\
\bottomrule
\end{tabular}

%% file: figures/fetch_data.tex
\newcommand{\fetchCachedOverheadMedianPerc}{6.50078573288968}
\newcommand{\fetchCachedOverheadMaxPerc}{10.4772879602681}
\newcommand{\fetchColdOverheadMin}{7.30647540663289}
\newcommand{\fetchColdOverheadMax}{422.388023726704}
\newcommand{\fetchNoCacheOverheadMin}{6.70864747455904}
\newcommand{\fetchNoCacheOverheadMax}{310.453168304745}
\newcommand{\fetchDiasporaSpreeModifiedOverheadMedianPerc}{2.25749195305619}
\newcommand{\fetchDiasporaSpreeModifiedOverheadMaxPerc}{4.59306183132621}
\newcommand{\fetchAutolabModifiedOverheadMedianPerc}{8.03726846984426}
\newcommand{\fetchAutolabModifiedOverheadMaxPerc}{21.0190676558494}
\newcommand{\fetchSpreeTwoModifiedFasterPerc}{10.736344761846}

%% file: figures/fetch.tex
\begin{tikzpicture}[x=1pt,y=1pt]
\definecolor{fillColor}{RGB}{255,255,255}
\path[use as bounding box,fill=fillColor,fill opacity=0.00] (0,0) rectangle (507.79,142.26);
\begin{scope}
\path[clip] (  0.00,  0.00) rectangle (507.79,142.26);
\definecolor{drawColor}{RGB}{255,255,255}
\definecolor{fillColor}{RGB}{255,255,255}

\path[draw=drawColor,line width= 0.4pt,line join=round,line cap=round,fill=fillColor] ( -0.00, -0.00) rectangle (507.79,142.26);
\end{scope}
\begin{scope}
\path[clip] ( 35.19, 36.37) rectangle (214.74,122.92);
\definecolor{fillColor}{RGB}{255,255,255}

\path[fill=fillColor] ( 35.19, 36.37) rectangle (214.74,122.92);
\definecolor{drawColor}{gray}{0.92}

\path[draw=drawColor,line width= 0.4pt,line join=round] ( 35.19, 36.41) --
	(214.74, 36.41);

\path[draw=drawColor,line width= 0.4pt,line join=round] ( 35.19, 57.83) --
	(214.74, 57.83);

\path[draw=drawColor,line width= 0.4pt,line join=round] ( 35.19, 79.25) --
	(214.74, 79.25);

\path[draw=drawColor,line width= 0.4pt,line join=round] ( 35.19,100.67) --
	(214.74,100.67);

\path[draw=drawColor,line width= 0.4pt,line join=round] ( 35.19,122.09) --
	(214.74,122.09);

\path[draw=drawColor,line width= 0.4pt,line join=round] ( 46.90, 36.37) --
	( 46.90,122.92);

\path[draw=drawColor,line width= 0.4pt,line join=round] ( 66.42, 36.37) --
	( 66.42,122.92);

\path[draw=drawColor,line width= 0.4pt,line join=round] ( 85.93, 36.37) --
	( 85.93,122.92);

\path[draw=drawColor,line width= 0.4pt,line join=round] (105.45, 36.37) --
	(105.45,122.92);

\path[draw=drawColor,line width= 0.4pt,line join=round] (124.97, 36.37) --
	(124.97,122.92);

\path[draw=drawColor,line width= 0.4pt,line join=round] (144.48, 36.37) --
	(144.48,122.92);

\path[draw=drawColor,line width= 0.4pt,line join=round] (164.00, 36.37) --
	(164.00,122.92);

\path[draw=drawColor,line width= 0.4pt,line join=round] (183.52, 36.37) --
	(183.52,122.92);

\path[draw=drawColor,line width= 0.4pt,line join=round] (203.03, 36.37) --
	(203.03,122.92);
\definecolor{fillColor}{RGB}{17,119,51}

\path[fill=fillColor] ( 44.94, 58.44) --
	( 48.86, 58.44) --
	( 48.86, 62.37) --
	( 44.94, 62.37) --
	cycle;
\definecolor{drawColor}{RGB}{170,68,153}

\path[draw=drawColor,line width= 0.4pt,line join=round,line cap=round] ( 46.08, 95.21) -- ( 51.63, 95.21);

\path[draw=drawColor,line width= 0.4pt,line join=round,line cap=round] ( 48.85, 92.43) -- ( 48.85, 97.98);
\definecolor{fillColor}{RGB}{204,102,119}

\path[fill=fillColor] ( 44.95, 63.19) --
	( 47.59, 58.61) --
	( 42.31, 58.61) --
	cycle;
\definecolor{drawColor}{RGB}{17,119,51}

\path[draw=drawColor,line width= 0.4pt,line join=round,line cap=round] ( 48.84, 85.72) rectangle ( 52.76, 89.64);

\path[draw=drawColor,line width= 0.4pt,line join=round,line cap=round] ( 48.84, 85.72) -- ( 52.76, 89.64);

\path[draw=drawColor,line width= 0.4pt,line join=round,line cap=round] ( 48.84, 89.64) -- ( 52.76, 85.72);
\definecolor{fillColor}{RGB}{51,34,136}

\path[fill=fillColor] ( 43.00, 59.88) circle (  1.96);
\definecolor{fillColor}{RGB}{17,119,51}

\path[fill=fillColor] ( 64.45, 46.42) --
	( 68.38, 46.42) --
	( 68.38, 50.35) --
	( 64.45, 50.35) --
	cycle;
\definecolor{drawColor}{RGB}{170,68,153}

\path[draw=drawColor,line width= 0.4pt,line join=round,line cap=round] ( 65.59, 80.67) -- ( 71.14, 80.67);

\path[draw=drawColor,line width= 0.4pt,line join=round,line cap=round] ( 68.37, 77.89) -- ( 68.37, 83.44);
\definecolor{fillColor}{RGB}{204,102,119}

\path[fill=fillColor] ( 64.46, 51.20) --
	( 67.11, 46.63) --
	( 61.82, 46.63) --
	cycle;
\definecolor{drawColor}{RGB}{17,119,51}

\path[draw=drawColor,line width= 0.4pt,line join=round,line cap=round] ( 68.36, 73.01) rectangle ( 72.28, 76.93);

\path[draw=drawColor,line width= 0.4pt,line join=round,line cap=round] ( 68.36, 73.01) -- ( 72.28, 76.93);

\path[draw=drawColor,line width= 0.4pt,line join=round,line cap=round] ( 68.36, 76.93) -- ( 72.28, 73.01);
\definecolor{fillColor}{RGB}{51,34,136}

\path[fill=fillColor] ( 62.51, 47.96) circle (  1.96);
\definecolor{fillColor}{RGB}{17,119,51}

\path[fill=fillColor] ( 83.97, 58.50) --
	( 87.90, 58.50) --
	( 87.90, 62.43) --
	( 83.97, 62.43) --
	cycle;
\definecolor{drawColor}{RGB}{170,68,153}

\path[draw=drawColor,line width= 0.4pt,line join=round,line cap=round] ( 85.11, 95.15) -- ( 90.66, 95.15);

\path[draw=drawColor,line width= 0.4pt,line join=round,line cap=round] ( 87.88, 92.37) -- ( 87.88, 97.92);
\definecolor{fillColor}{RGB}{204,102,119}

\path[fill=fillColor] ( 83.98, 63.25) --
	( 86.62, 58.67) --
	( 81.34, 58.67) --
	cycle;
\definecolor{drawColor}{RGB}{17,119,51}

\path[draw=drawColor,line width= 0.4pt,line join=round,line cap=round] ( 87.87, 86.07) rectangle ( 91.80, 90.00);

\path[draw=drawColor,line width= 0.4pt,line join=round,line cap=round] ( 87.87, 86.07) -- ( 91.80, 90.00);

\path[draw=drawColor,line width= 0.4pt,line join=round,line cap=round] ( 87.87, 90.00) -- ( 91.80, 86.07);
\definecolor{fillColor}{RGB}{51,34,136}

\path[fill=fillColor] ( 82.03, 59.99) circle (  1.96);
\definecolor{fillColor}{RGB}{17,119,51}

\path[fill=fillColor] (103.49, 61.04) --
	(107.41, 61.04) --
	(107.41, 64.96) --
	(103.49, 64.96) --
	cycle;
\definecolor{drawColor}{RGB}{170,68,153}

\path[draw=drawColor,line width= 0.4pt,line join=round,line cap=round] (104.63, 82.39) -- (110.18, 82.39);

\path[draw=drawColor,line width= 0.4pt,line join=round,line cap=round] (107.40, 79.62) -- (107.40, 85.17);
\definecolor{fillColor}{RGB}{204,102,119}

\path[fill=fillColor] (103.50, 65.75) --
	(106.14, 61.17) --
	(100.86, 61.17) --
	cycle;
\definecolor{drawColor}{RGB}{17,119,51}

\path[draw=drawColor,line width= 0.4pt,line join=round,line cap=round] (107.39, 95.69) rectangle (111.32, 99.62);

\path[draw=drawColor,line width= 0.4pt,line join=round,line cap=round] (107.39, 95.69) -- (111.32, 99.62);

\path[draw=drawColor,line width= 0.4pt,line join=round,line cap=round] (107.39, 99.62) -- (111.32, 95.69);
\definecolor{fillColor}{RGB}{51,34,136}

\path[fill=fillColor] (101.55, 62.57) circle (  1.96);
\definecolor{fillColor}{RGB}{17,119,51}

\path[fill=fillColor] (123.00, 43.03) --
	(126.93, 43.03) --
	(126.93, 46.95) --
	(123.00, 46.95) --
	cycle;
\definecolor{drawColor}{RGB}{170,68,153}

\path[draw=drawColor,line width= 0.4pt,line join=round,line cap=round] (124.14, 74.58) -- (129.69, 74.58);

\path[draw=drawColor,line width= 0.4pt,line join=round,line cap=round] (126.92, 71.80) -- (126.92, 77.35);
\definecolor{fillColor}{RGB}{204,102,119}

\path[fill=fillColor] (123.02, 47.84) --
	(125.66, 43.27) --
	(120.37, 43.27) --
	cycle;
\definecolor{drawColor}{RGB}{17,119,51}

\path[draw=drawColor,line width= 0.4pt,line join=round,line cap=round] (126.91, 63.97) rectangle (130.83, 67.89);

\path[draw=drawColor,line width= 0.4pt,line join=round,line cap=round] (126.91, 63.97) -- (130.83, 67.89);

\path[draw=drawColor,line width= 0.4pt,line join=round,line cap=round] (126.91, 67.89) -- (130.83, 63.97);
\definecolor{fillColor}{RGB}{51,34,136}

\path[fill=fillColor] (121.06, 44.67) circle (  1.96);
\definecolor{fillColor}{RGB}{17,119,51}

\path[fill=fillColor] (142.52, 62.86) --
	(146.45, 62.86) --
	(146.45, 66.78) --
	(142.52, 66.78) --
	cycle;
\definecolor{drawColor}{RGB}{170,68,153}

\path[draw=drawColor,line width= 0.4pt,line join=round,line cap=round] (143.66, 93.17) -- (149.21, 93.17);

\path[draw=drawColor,line width= 0.4pt,line join=round,line cap=round] (146.44, 90.40) -- (146.44, 95.95);
\definecolor{fillColor}{RGB}{204,102,119}

\path[fill=fillColor] (142.53, 67.69) --
	(145.17, 63.11) --
	(139.89, 63.11) --
	cycle;
\definecolor{drawColor}{RGB}{17,119,51}

\path[draw=drawColor,line width= 0.4pt,line join=round,line cap=round] (146.42, 83.46) rectangle (150.35, 87.38);

\path[draw=drawColor,line width= 0.4pt,line join=round,line cap=round] (146.42, 83.46) -- (150.35, 87.38);

\path[draw=drawColor,line width= 0.4pt,line join=round,line cap=round] (146.42, 87.38) -- (150.35, 83.46);
\definecolor{fillColor}{RGB}{51,34,136}

\path[fill=fillColor] (140.58, 64.41) circle (  1.96);
\definecolor{fillColor}{RGB}{17,119,51}

\path[fill=fillColor] (162.04, 56.39) --
	(165.96, 56.39) --
	(165.96, 60.32) --
	(162.04, 60.32) --
	cycle;
\definecolor{drawColor}{RGB}{170,68,153}

\path[draw=drawColor,line width= 0.4pt,line join=round,line cap=round] (163.18, 88.65) -- (168.73, 88.65);

\path[draw=drawColor,line width= 0.4pt,line join=round,line cap=round] (165.95, 85.87) -- (165.95, 91.42);
\definecolor{fillColor}{RGB}{204,102,119}

\path[fill=fillColor] (162.05, 61.06) --
	(164.69, 56.49) --
	(159.41, 56.49) --
	cycle;
\definecolor{drawColor}{RGB}{17,119,51}

\path[draw=drawColor,line width= 0.4pt,line join=round,line cap=round] (165.94, 79.20) rectangle (169.87, 83.13);

\path[draw=drawColor,line width= 0.4pt,line join=round,line cap=round] (165.94, 79.20) -- (169.87, 83.13);

\path[draw=drawColor,line width= 0.4pt,line join=round,line cap=round] (165.94, 83.13) -- (169.87, 79.20);
\definecolor{fillColor}{RGB}{51,34,136}

\path[fill=fillColor] (160.10, 57.61) circle (  1.96);
\definecolor{fillColor}{RGB}{17,119,51}

\path[fill=fillColor] (181.56, 51.38) --
	(185.48, 51.38) --
	(185.48, 55.31) --
	(181.56, 55.31) --
	cycle;
\definecolor{drawColor}{RGB}{170,68,153}

\path[draw=drawColor,line width= 0.4pt,line join=round,line cap=round] (182.69, 90.27) -- (188.24, 90.27);

\path[draw=drawColor,line width= 0.4pt,line join=round,line cap=round] (185.47, 87.49) -- (185.47, 93.04);
\definecolor{fillColor}{RGB}{204,102,119}

\path[fill=fillColor] (181.57, 55.69) --
	(184.21, 51.11) --
	(178.92, 51.11) --
	cycle;
\definecolor{drawColor}{RGB}{17,119,51}

\path[draw=drawColor,line width= 0.4pt,line join=round,line cap=round] (185.46, 86.40) rectangle (189.38, 90.32);

\path[draw=drawColor,line width= 0.4pt,line join=round,line cap=round] (185.46, 86.40) -- (189.38, 90.32);

\path[draw=drawColor,line width= 0.4pt,line join=round,line cap=round] (185.46, 90.32) -- (189.38, 86.40);
\definecolor{fillColor}{RGB}{51,34,136}

\path[fill=fillColor] (179.61, 52.50) circle (  1.96);
\definecolor{fillColor}{RGB}{17,119,51}

\path[fill=fillColor] (201.07, 47.09) --
	(205.00, 47.09) --
	(205.00, 51.01) --
	(201.07, 51.01) --
	cycle;
\definecolor{drawColor}{RGB}{170,68,153}

\path[draw=drawColor,line width= 0.4pt,line join=round,line cap=round] (202.21, 85.49) -- (207.76, 85.49);

\path[draw=drawColor,line width= 0.4pt,line join=round,line cap=round] (204.99, 82.71) -- (204.99, 88.26);
\definecolor{fillColor}{RGB}{204,102,119}

\path[fill=fillColor] (201.08, 51.79) --
	(203.73, 47.21) --
	(198.44, 47.21) --
	cycle;
\definecolor{drawColor}{RGB}{17,119,51}

\path[draw=drawColor,line width= 0.4pt,line join=round,line cap=round] (204.98, 75.31) rectangle (208.90, 79.23);

\path[draw=drawColor,line width= 0.4pt,line join=round,line cap=round] (204.98, 75.31) -- (208.90, 79.23);

\path[draw=drawColor,line width= 0.4pt,line join=round,line cap=round] (204.98, 79.23) -- (208.90, 75.31);
\definecolor{fillColor}{RGB}{51,34,136}

\path[fill=fillColor] (199.13, 48.48) circle (  1.96);
\definecolor{drawColor}{gray}{0.20}

\path[draw=drawColor,line width= 0.4pt,line join=round,line cap=round] ( 35.19, 36.37) rectangle (214.74,122.92);
\end{scope}
\begin{scope}
\path[clip] (218.74, 36.37) rectangle (378.78,122.92);
\definecolor{fillColor}{RGB}{255,255,255}

\path[fill=fillColor] (218.74, 36.37) rectangle (378.78,122.92);
\definecolor{drawColor}{gray}{0.92}

\path[draw=drawColor,line width= 0.4pt,line join=round] (218.74, 36.41) --
	(378.78, 36.41);

\path[draw=drawColor,line width= 0.4pt,line join=round] (218.74, 57.83) --
	(378.78, 57.83);

\path[draw=drawColor,line width= 0.4pt,line join=round] (218.74, 79.25) --
	(378.78, 79.25);

\path[draw=drawColor,line width= 0.4pt,line join=round] (218.74,100.67) --
	(378.78,100.67);

\path[draw=drawColor,line width= 0.4pt,line join=round] (218.74,122.09) --
	(378.78,122.09);

\path[draw=drawColor,line width= 0.4pt,line join=round] (230.45, 36.37) --
	(230.45,122.92);

\path[draw=drawColor,line width= 0.4pt,line join=round] (249.97, 36.37) --
	(249.97,122.92);

\path[draw=drawColor,line width= 0.4pt,line join=round] (269.49, 36.37) --
	(269.49,122.92);

\path[draw=drawColor,line width= 0.4pt,line join=round] (289.00, 36.37) --
	(289.00,122.92);

\path[draw=drawColor,line width= 0.4pt,line join=round] (308.52, 36.37) --
	(308.52,122.92);

\path[draw=drawColor,line width= 0.4pt,line join=round] (328.04, 36.37) --
	(328.04,122.92);

\path[draw=drawColor,line width= 0.4pt,line join=round] (347.56, 36.37) --
	(347.56,122.92);

\path[draw=drawColor,line width= 0.4pt,line join=round] (367.07, 36.37) --
	(367.07,122.92);
\definecolor{fillColor}{RGB}{17,119,51}

\path[fill=fillColor] (228.49, 50.09) --
	(232.42, 50.09) --
	(232.42, 54.01) --
	(228.49, 54.01) --
	cycle;
\definecolor{drawColor}{RGB}{170,68,153}

\path[draw=drawColor,line width= 0.4pt,line join=round,line cap=round] (229.63, 81.35) -- (235.18, 81.35);

\path[draw=drawColor,line width= 0.4pt,line join=round,line cap=round] (232.41, 78.58) -- (232.41, 84.13);
\definecolor{fillColor}{RGB}{204,102,119}

\path[fill=fillColor] (228.50, 54.52) --
	(231.15, 49.94) --
	(225.86, 49.94) --
	cycle;
\definecolor{drawColor}{RGB}{17,119,51}

\path[draw=drawColor,line width= 0.4pt,line join=round,line cap=round] (232.40, 71.65) rectangle (236.32, 75.58);

\path[draw=drawColor,line width= 0.4pt,line join=round,line cap=round] (232.40, 71.65) -- (236.32, 75.58);

\path[draw=drawColor,line width= 0.4pt,line join=round,line cap=round] (232.40, 75.58) -- (236.32, 71.65);
\definecolor{fillColor}{RGB}{51,34,136}

\path[fill=fillColor] (226.55, 51.33) circle (  1.96);
\definecolor{fillColor}{RGB}{17,119,51}

\path[fill=fillColor] (248.01, 53.47) --
	(251.93, 53.47) --
	(251.93, 57.39) --
	(248.01, 57.39) --
	cycle;
\definecolor{drawColor}{RGB}{170,68,153}

\path[draw=drawColor,line width= 0.4pt,line join=round,line cap=round] (249.15,101.01) -- (254.70,101.01);

\path[draw=drawColor,line width= 0.4pt,line join=round,line cap=round] (251.92, 98.24) -- (251.92,103.79);
\definecolor{fillColor}{RGB}{204,102,119}

\path[fill=fillColor] (248.02, 57.57) --
	(250.66, 53.00) --
	(245.38, 53.00) --
	cycle;
\definecolor{drawColor}{RGB}{17,119,51}

\path[draw=drawColor,line width= 0.4pt,line join=round,line cap=round] (251.91, 90.80) rectangle (255.84, 94.72);

\path[draw=drawColor,line width= 0.4pt,line join=round,line cap=round] (251.91, 90.80) -- (255.84, 94.72);

\path[draw=drawColor,line width= 0.4pt,line join=round,line cap=round] (251.91, 94.72) -- (255.84, 90.80);
\definecolor{fillColor}{RGB}{51,34,136}

\path[fill=fillColor] (246.07, 55.58) circle (  1.96);
\definecolor{fillColor}{RGB}{17,119,51}

\path[fill=fillColor] (267.53, 41.44) --
	(271.45, 41.44) --
	(271.45, 45.36) --
	(267.53, 45.36) --
	cycle;
\definecolor{drawColor}{RGB}{170,68,153}

\path[draw=drawColor,line width= 0.4pt,line join=round,line cap=round] (268.66, 77.31) -- (274.21, 77.31);

\path[draw=drawColor,line width= 0.4pt,line join=round,line cap=round] (271.44, 74.53) -- (271.44, 80.08);
\definecolor{fillColor}{RGB}{204,102,119}

\path[fill=fillColor] (267.54, 45.96) --
	(270.18, 41.38) --
	(264.89, 41.38) --
	cycle;
\definecolor{drawColor}{RGB}{17,119,51}

\path[draw=drawColor,line width= 0.4pt,line join=round,line cap=round] (271.43, 67.10) rectangle (275.35, 71.03);

\path[draw=drawColor,line width= 0.4pt,line join=round,line cap=round] (271.43, 67.10) -- (275.35, 71.03);

\path[draw=drawColor,line width= 0.4pt,line join=round,line cap=round] (271.43, 71.03) -- (275.35, 67.10);
\definecolor{fillColor}{RGB}{51,34,136}

\path[fill=fillColor] (265.58, 42.54) circle (  1.96);
\definecolor{fillColor}{RGB}{17,119,51}

\path[fill=fillColor] (287.04, 54.40) --
	(290.97, 54.40) --
	(290.97, 58.32) --
	(287.04, 58.32) --
	cycle;
\definecolor{drawColor}{RGB}{170,68,153}

\path[draw=drawColor,line width= 0.4pt,line join=round,line cap=round] (288.18,104.80) -- (293.73,104.80);

\path[draw=drawColor,line width= 0.4pt,line join=round,line cap=round] (290.96,102.03) -- (290.96,107.58);
\definecolor{fillColor}{RGB}{204,102,119}

\path[fill=fillColor] (287.05, 58.62) --
	(289.70, 54.04) --
	(284.41, 54.04) --
	cycle;
\definecolor{drawColor}{RGB}{17,119,51}

\path[draw=drawColor,line width= 0.4pt,line join=round,line cap=round] (290.95, 96.06) rectangle (294.87, 99.99);

\path[draw=drawColor,line width= 0.4pt,line join=round,line cap=round] (290.95, 96.06) -- (294.87, 99.99);

\path[draw=drawColor,line width= 0.4pt,line join=round,line cap=round] (290.95, 99.99) -- (294.87, 96.06);
\definecolor{fillColor}{RGB}{51,34,136}

\path[fill=fillColor] (285.10, 55.40) circle (  1.96);
\definecolor{fillColor}{RGB}{17,119,51}

\path[fill=fillColor] (306.56, 58.72) --
	(310.48, 58.72) --
	(310.48, 62.64) --
	(306.56, 62.64) --
	cycle;
\definecolor{drawColor}{RGB}{170,68,153}

\path[draw=drawColor,line width= 0.4pt,line join=round,line cap=round] (307.70,115.90) -- (313.25,115.90);

\path[draw=drawColor,line width= 0.4pt,line join=round,line cap=round] (310.47,113.12) -- (310.47,118.67);
\definecolor{fillColor}{RGB}{204,102,119}

\path[fill=fillColor] (306.57, 62.95) --
	(309.21, 58.37) --
	(303.93, 58.37) --
	cycle;
\definecolor{drawColor}{RGB}{17,119,51}

\path[draw=drawColor,line width= 0.4pt,line join=round,line cap=round] (310.46,111.34) rectangle (314.39,115.27);

\path[draw=drawColor,line width= 0.4pt,line join=round,line cap=round] (310.46,111.34) -- (314.39,115.27);

\path[draw=drawColor,line width= 0.4pt,line join=round,line cap=round] (310.46,115.27) -- (314.39,111.34);
\definecolor{fillColor}{RGB}{51,34,136}

\path[fill=fillColor] (304.62, 59.85) circle (  1.96);
\definecolor{fillColor}{RGB}{17,119,51}

\path[fill=fillColor] (326.08, 39.17) --
	(330.00, 39.17) --
	(330.00, 43.09) --
	(326.08, 43.09) --
	cycle;
\definecolor{drawColor}{RGB}{170,68,153}

\path[draw=drawColor,line width= 0.4pt,line join=round,line cap=round] (327.22, 68.12) -- (332.77, 68.12);

\path[draw=drawColor,line width= 0.4pt,line join=round,line cap=round] (329.99, 65.34) -- (329.99, 70.89);
\definecolor{fillColor}{RGB}{204,102,119}

\path[fill=fillColor] (326.09, 43.64) --
	(328.73, 39.07) --
	(323.44, 39.07) --
	cycle;
\definecolor{drawColor}{RGB}{17,119,51}

\path[draw=drawColor,line width= 0.4pt,line join=round,line cap=round] (329.98, 57.63) rectangle (333.90, 61.55);

\path[draw=drawColor,line width= 0.4pt,line join=round,line cap=round] (329.98, 57.63) -- (333.90, 61.55);

\path[draw=drawColor,line width= 0.4pt,line join=round,line cap=round] (329.98, 61.55) -- (333.90, 57.63);
\definecolor{fillColor}{RGB}{51,34,136}

\path[fill=fillColor] (324.14, 40.30) circle (  1.96);
\definecolor{fillColor}{RGB}{17,119,51}

\path[fill=fillColor] (345.59, 39.52) --
	(349.52, 39.52) --
	(349.52, 43.45) --
	(345.59, 43.45) --
	cycle;
\definecolor{drawColor}{RGB}{170,68,153}

\path[draw=drawColor,line width= 0.4pt,line join=round,line cap=round] (346.73, 72.75) -- (352.28, 72.75);

\path[draw=drawColor,line width= 0.4pt,line join=round,line cap=round] (349.51, 69.97) -- (349.51, 75.52);
\definecolor{fillColor}{RGB}{204,102,119}

\path[fill=fillColor] (345.60, 43.93) --
	(348.25, 39.35) --
	(342.96, 39.35) --
	cycle;
\definecolor{drawColor}{RGB}{17,119,51}

\path[draw=drawColor,line width= 0.4pt,line join=round,line cap=round] (349.50, 63.46) rectangle (353.42, 67.39);

\path[draw=drawColor,line width= 0.4pt,line join=round,line cap=round] (349.50, 63.46) -- (353.42, 67.39);

\path[draw=drawColor,line width= 0.4pt,line join=round,line cap=round] (349.50, 67.39) -- (353.42, 63.46);
\definecolor{fillColor}{RGB}{51,34,136}

\path[fill=fillColor] (343.65, 40.46) circle (  1.96);
\definecolor{fillColor}{RGB}{17,119,51}

\path[fill=fillColor] (365.11, 39.53) --
	(369.03, 39.53) --
	(369.03, 43.46) --
	(365.11, 43.46) --
	cycle;
\definecolor{drawColor}{RGB}{170,68,153}

\path[draw=drawColor,line width= 0.4pt,line join=round,line cap=round] (366.25, 72.88) -- (371.80, 72.88);

\path[draw=drawColor,line width= 0.4pt,line join=round,line cap=round] (369.02, 70.10) -- (369.02, 75.65);
\definecolor{fillColor}{RGB}{204,102,119}

\path[fill=fillColor] (365.12, 43.99) --
	(367.76, 39.42) --
	(362.48, 39.42) --
	cycle;
\definecolor{drawColor}{RGB}{17,119,51}

\path[draw=drawColor,line width= 0.4pt,line join=round,line cap=round] (369.01, 63.55) rectangle (372.94, 67.48);

\path[draw=drawColor,line width= 0.4pt,line join=round,line cap=round] (369.01, 63.55) -- (372.94, 67.48);

\path[draw=drawColor,line width= 0.4pt,line join=round,line cap=round] (369.01, 67.48) -- (372.94, 63.55);
\definecolor{fillColor}{RGB}{51,34,136}

\path[fill=fillColor] (363.17, 40.54) circle (  1.96);
\definecolor{drawColor}{gray}{0.20}

\path[draw=drawColor,line width= 0.4pt,line join=round,line cap=round] (218.74, 36.37) rectangle (378.78,122.92);
\end{scope}
\begin{scope}
\path[clip] (382.78, 36.37) rectangle (503.79,122.92);
\definecolor{fillColor}{RGB}{255,255,255}

\path[fill=fillColor] (382.78, 36.37) rectangle (503.79,122.92);
\definecolor{drawColor}{gray}{0.92}

\path[draw=drawColor,line width= 0.4pt,line join=round] (382.78, 36.41) --
	(503.79, 36.41);

\path[draw=drawColor,line width= 0.4pt,line join=round] (382.78, 57.83) --
	(503.79, 57.83);

\path[draw=drawColor,line width= 0.4pt,line join=round] (382.78, 79.25) --
	(503.79, 79.25);

\path[draw=drawColor,line width= 0.4pt,line join=round] (382.78,100.67) --
	(503.79,100.67);

\path[draw=drawColor,line width= 0.4pt,line join=round] (382.78,122.09) --
	(503.79,122.09);

\path[draw=drawColor,line width= 0.4pt,line join=round] (394.49, 36.37) --
	(394.49,122.92);

\path[draw=drawColor,line width= 0.4pt,line join=round] (414.01, 36.37) --
	(414.01,122.92);

\path[draw=drawColor,line width= 0.4pt,line join=round] (433.53, 36.37) --
	(433.53,122.92);

\path[draw=drawColor,line width= 0.4pt,line join=round] (453.04, 36.37) --
	(453.04,122.92);

\path[draw=drawColor,line width= 0.4pt,line join=round] (472.56, 36.37) --
	(472.56,122.92);

\path[draw=drawColor,line width= 0.4pt,line join=round] (492.08, 36.37) --
	(492.08,122.92);
\definecolor{fillColor}{RGB}{17,119,51}

\path[fill=fillColor] (392.53, 44.77) --
	(396.45, 44.77) --
	(396.45, 48.70) --
	(392.53, 48.70) --
	cycle;
\definecolor{drawColor}{RGB}{170,68,153}

\path[draw=drawColor,line width= 0.4pt,line join=round,line cap=round] (393.67, 83.10) -- (399.22, 83.10);

\path[draw=drawColor,line width= 0.4pt,line join=round,line cap=round] (396.44, 80.33) -- (396.44, 85.88);
\definecolor{fillColor}{RGB}{204,102,119}

\path[fill=fillColor] (392.54, 49.13) --
	(395.18, 44.55) --
	(389.90, 44.55) --
	cycle;
\definecolor{drawColor}{RGB}{17,119,51}

\path[draw=drawColor,line width= 0.4pt,line join=round,line cap=round] (396.43, 80.41) rectangle (400.36, 84.34);

\path[draw=drawColor,line width= 0.4pt,line join=round,line cap=round] (396.43, 80.41) -- (400.36, 84.34);

\path[draw=drawColor,line width= 0.4pt,line join=round,line cap=round] (396.43, 84.34) -- (400.36, 80.41);
\definecolor{fillColor}{RGB}{51,34,136}

\path[fill=fillColor] (390.59, 44.94) circle (  1.96);
\definecolor{fillColor}{RGB}{17,119,51}

\path[fill=fillColor] (412.05, 45.89) --
	(415.97, 45.89) --
	(415.97, 49.82) --
	(412.05, 49.82) --
	cycle;
\definecolor{drawColor}{RGB}{170,68,153}

\path[draw=drawColor,line width= 0.4pt,line join=round,line cap=round] (413.19, 96.47) -- (418.74, 96.47);

\path[draw=drawColor,line width= 0.4pt,line join=round,line cap=round] (415.96, 93.69) -- (415.96, 99.24);
\definecolor{fillColor}{RGB}{204,102,119}

\path[fill=fillColor] (412.06, 49.99) --
	(414.70, 45.41) --
	(409.42, 45.41) --
	cycle;
\definecolor{drawColor}{RGB}{17,119,51}

\path[draw=drawColor,line width= 0.4pt,line join=round,line cap=round] (415.95, 89.84) rectangle (419.88, 93.76);

\path[draw=drawColor,line width= 0.4pt,line join=round,line cap=round] (415.95, 89.84) -- (419.88, 93.76);

\path[draw=drawColor,line width= 0.4pt,line join=round,line cap=round] (415.95, 93.76) -- (419.88, 89.84);
\definecolor{fillColor}{RGB}{51,34,136}

\path[fill=fillColor] (410.11, 46.52) circle (  1.96);
\definecolor{fillColor}{RGB}{17,119,51}

\path[fill=fillColor] (431.56, 39.43) --
	(435.49, 39.43) --
	(435.49, 43.36) --
	(431.56, 43.36) --
	cycle;
\definecolor{drawColor}{RGB}{170,68,153}

\path[draw=drawColor,line width= 0.4pt,line join=round,line cap=round] (432.70, 78.72) -- (438.25, 78.72);

\path[draw=drawColor,line width= 0.4pt,line join=round,line cap=round] (435.48, 75.95) -- (435.48, 81.50);
\definecolor{fillColor}{RGB}{204,102,119}

\path[fill=fillColor] (431.57, 43.83) --
	(434.22, 39.25) --
	(428.93, 39.25) --
	cycle;
\definecolor{drawColor}{RGB}{17,119,51}

\path[draw=drawColor,line width= 0.4pt,line join=round,line cap=round] (435.47, 69.42) rectangle (439.39, 73.35);

\path[draw=drawColor,line width= 0.4pt,line join=round,line cap=round] (435.47, 69.42) -- (439.39, 73.35);

\path[draw=drawColor,line width= 0.4pt,line join=round,line cap=round] (435.47, 73.35) -- (439.39, 69.42);
\definecolor{fillColor}{RGB}{51,34,136}

\path[fill=fillColor] (429.62, 40.36) circle (  1.96);
\definecolor{fillColor}{RGB}{17,119,51}

\path[fill=fillColor] (451.08, 48.12) --
	(455.01, 48.12) --
	(455.01, 52.05) --
	(451.08, 52.05) --
	cycle;
\definecolor{drawColor}{RGB}{170,68,153}

\path[draw=drawColor,line width= 0.4pt,line join=round,line cap=round] (452.22,102.66) -- (457.77,102.66);

\path[draw=drawColor,line width= 0.4pt,line join=round,line cap=round] (454.99, 99.88) -- (454.99,105.43);
\definecolor{fillColor}{RGB}{204,102,119}

\path[fill=fillColor] (451.09, 52.21) --
	(453.73, 47.63) --
	(448.45, 47.63) --
	cycle;
\definecolor{drawColor}{RGB}{17,119,51}

\path[draw=drawColor,line width= 0.4pt,line join=round,line cap=round] (454.98, 88.82) rectangle (458.91, 92.74);

\path[draw=drawColor,line width= 0.4pt,line join=round,line cap=round] (454.98, 88.82) -- (458.91, 92.74);

\path[draw=drawColor,line width= 0.4pt,line join=round,line cap=round] (454.98, 92.74) -- (458.91, 88.82);
\definecolor{fillColor}{RGB}{51,34,136}

\path[fill=fillColor] (449.14, 48.17) circle (  1.96);
\definecolor{fillColor}{RGB}{17,119,51}

\path[fill=fillColor] (470.60, 42.34) --
	(474.52, 42.34) --
	(474.52, 46.26) --
	(470.60, 46.26) --
	cycle;
\definecolor{drawColor}{RGB}{170,68,153}

\path[draw=drawColor,line width= 0.4pt,line join=round,line cap=round] (471.74, 99.77) -- (477.29, 99.77);

\path[draw=drawColor,line width= 0.4pt,line join=round,line cap=round] (474.51, 97.00) -- (474.51,102.55);
\definecolor{fillColor}{RGB}{204,102,119}

\path[fill=fillColor] (470.61, 46.56) --
	(473.25, 41.98) --
	(467.97, 41.98) --
	cycle;
\definecolor{drawColor}{RGB}{17,119,51}

\path[draw=drawColor,line width= 0.4pt,line join=round,line cap=round] (474.50, 78.07) rectangle (478.43, 81.99);

\path[draw=drawColor,line width= 0.4pt,line join=round,line cap=round] (474.50, 78.07) -- (478.43, 81.99);

\path[draw=drawColor,line width= 0.4pt,line join=round,line cap=round] (474.50, 81.99) -- (478.43, 78.07);
\definecolor{fillColor}{RGB}{51,34,136}

\path[fill=fillColor] (468.66, 41.73) circle (  1.96);
\definecolor{fillColor}{RGB}{17,119,51}

\path[fill=fillColor] (490.11, 66.78) --
	(494.04, 66.78) --
	(494.04, 70.70) --
	(490.11, 70.70) --
	cycle;
\definecolor{drawColor}{RGB}{170,68,153}

\path[draw=drawColor,line width= 0.4pt,line join=round,line cap=round] (491.25,102.22) -- (496.80,102.22);

\path[draw=drawColor,line width= 0.4pt,line join=round,line cap=round] (494.03, 99.44) -- (494.03,104.99);
\definecolor{fillColor}{RGB}{204,102,119}

\path[fill=fillColor] (490.13, 70.99) --
	(492.77, 66.41) --
	(487.48, 66.41) --
	cycle;
\definecolor{drawColor}{RGB}{17,119,51}

\path[draw=drawColor,line width= 0.4pt,line join=round,line cap=round] (494.02,117.03) rectangle (497.94,120.95);

\path[draw=drawColor,line width= 0.4pt,line join=round,line cap=round] (494.02,117.03) -- (497.94,120.95);

\path[draw=drawColor,line width= 0.4pt,line join=round,line cap=round] (494.02,120.95) -- (497.94,117.03);
\definecolor{fillColor}{RGB}{51,34,136}

\path[fill=fillColor] (488.17, 67.49) circle (  1.96);
\definecolor{drawColor}{gray}{0.20}

\path[draw=drawColor,line width= 0.4pt,line join=round,line cap=round] (382.78, 36.37) rectangle (503.79,122.92);
\end{scope}
\begin{scope}
\path[clip] ( 35.19, 13.06) rectangle (214.74, 25.12);
\definecolor{drawColor}{gray}{0.20}
\definecolor{fillColor}{gray}{0.85}

\path[draw=drawColor,line width= 0.4pt,line join=round,line cap=round,fill=fillColor] ( 35.19, 13.06) rectangle (214.74, 25.12);
\definecolor{drawColor}{gray}{0.10}

\node[text=drawColor,anchor=base,inner sep=0pt, outer sep=0pt, scale=  0.64] at (124.97, 16.89) {diaspora*};
\end{scope}
\begin{scope}
\path[clip] (218.74, 13.06) rectangle (378.78, 25.12);
\definecolor{drawColor}{gray}{0.20}
\definecolor{fillColor}{gray}{0.85}

\path[draw=drawColor,line width= 0.4pt,line join=round,line cap=round,fill=fillColor] (218.74, 13.06) rectangle (378.78, 25.12);
\definecolor{drawColor}{gray}{0.10}

\node[text=drawColor,anchor=base,inner sep=0pt, outer sep=0pt, scale=  0.64] at (298.76, 16.89) {Spree};
\end{scope}
\begin{scope}
\path[clip] (382.78, 13.06) rectangle (503.79, 25.12);
\definecolor{drawColor}{gray}{0.20}
\definecolor{fillColor}{gray}{0.85}

\path[draw=drawColor,line width= 0.4pt,line join=round,line cap=round,fill=fillColor] (382.78, 13.06) rectangle (503.79, 25.12);
\definecolor{drawColor}{gray}{0.10}

\node[text=drawColor,anchor=base,inner sep=0pt, outer sep=0pt, scale=  0.64] at (443.28, 16.89) {Autolab};
\end{scope}
\begin{scope}
\path[clip] (  0.00,  0.00) rectangle (507.79,142.26);
\definecolor{drawColor}{gray}{0.20}

\path[draw=drawColor,line width= 0.4pt,line join=round] ( 46.90, 34.37) --
	( 46.90, 36.37);

\path[draw=drawColor,line width= 0.4pt,line join=round] ( 66.42, 34.37) --
	( 66.42, 36.37);

\path[draw=drawColor,line width= 0.4pt,line join=round] ( 85.93, 34.37) --
	( 85.93, 36.37);

\path[draw=drawColor,line width= 0.4pt,line join=round] (105.45, 34.37) --
	(105.45, 36.37);

\path[draw=drawColor,line width= 0.4pt,line join=round] (124.97, 34.37) --
	(124.97, 36.37);

\path[draw=drawColor,line width= 0.4pt,line join=round] (144.48, 34.37) --
	(144.48, 36.37);

\path[draw=drawColor,line width= 0.4pt,line join=round] (164.00, 34.37) --
	(164.00, 36.37);

\path[draw=drawColor,line width= 0.4pt,line join=round] (183.52, 34.37) --
	(183.52, 36.37);

\path[draw=drawColor,line width= 0.4pt,line join=round] (203.03, 34.37) --
	(203.03, 36.37);
\end{scope}
\begin{scope}
\path[clip] (  0.00,  0.00) rectangle (507.79,142.26);
\definecolor{drawColor}{gray}{0.30}

\node[text=drawColor,anchor=base,inner sep=0pt, outer sep=0pt, scale=  0.64] at ( 46.90, 28.36) {D1};

\node[text=drawColor,anchor=base,inner sep=0pt, outer sep=0pt, scale=  0.64] at ( 66.42, 28.36) {D2};

\node[text=drawColor,anchor=base,inner sep=0pt, outer sep=0pt, scale=  0.64] at ( 85.93, 28.36) {D3};

\node[text=drawColor,anchor=base,inner sep=0pt, outer sep=0pt, scale=  0.64] at (105.45, 28.36) {D4};

\node[text=drawColor,anchor=base,inner sep=0pt, outer sep=0pt, scale=  0.64] at (124.97, 28.36) {D5};

\node[text=drawColor,anchor=base,inner sep=0pt, outer sep=0pt, scale=  0.64] at (144.48, 28.36) {D6};

\node[text=drawColor,anchor=base,inner sep=0pt, outer sep=0pt, scale=  0.64] at (164.00, 28.36) {D7};

\node[text=drawColor,anchor=base,inner sep=0pt, outer sep=0pt, scale=  0.64] at (183.52, 28.36) {D8};

\node[text=drawColor,anchor=base,inner sep=0pt, outer sep=0pt, scale=  0.64] at (203.03, 28.36) {D9};
\end{scope}
\begin{scope}
\path[clip] (  0.00,  0.00) rectangle (507.79,142.26);
\definecolor{drawColor}{gray}{0.20}

\path[draw=drawColor,line width= 0.4pt,line join=round] (230.45, 34.37) --
	(230.45, 36.37);

\path[draw=drawColor,line width= 0.4pt,line join=round] (249.97, 34.37) --
	(249.97, 36.37);

\path[draw=drawColor,line width= 0.4pt,line join=round] (269.49, 34.37) --
	(269.49, 36.37);

\path[draw=drawColor,line width= 0.4pt,line join=round] (289.00, 34.37) --
	(289.00, 36.37);

\path[draw=drawColor,line width= 0.4pt,line join=round] (308.52, 34.37) --
	(308.52, 36.37);

\path[draw=drawColor,line width= 0.4pt,line join=round] (328.04, 34.37) --
	(328.04, 36.37);

\path[draw=drawColor,line width= 0.4pt,line join=round] (347.56, 34.37) --
	(347.56, 36.37);

\path[draw=drawColor,line width= 0.4pt,line join=round] (367.07, 34.37) --
	(367.07, 36.37);
\end{scope}
\begin{scope}
\path[clip] (  0.00,  0.00) rectangle (507.79,142.26);
\definecolor{drawColor}{gray}{0.30}

\node[text=drawColor,anchor=base,inner sep=0pt, outer sep=0pt, scale=  0.64] at (230.45, 28.36) {S1};

\node[text=drawColor,anchor=base,inner sep=0pt, outer sep=0pt, scale=  0.64] at (249.97, 28.36) {S2};

\node[text=drawColor,anchor=base,inner sep=0pt, outer sep=0pt, scale=  0.64] at (269.49, 28.36) {S3};

\node[text=drawColor,anchor=base,inner sep=0pt, outer sep=0pt, scale=  0.64] at (289.00, 28.36) {S4};

\node[text=drawColor,anchor=base,inner sep=0pt, outer sep=0pt, scale=  0.64] at (308.52, 28.36) {S5};

\node[text=drawColor,anchor=base,inner sep=0pt, outer sep=0pt, scale=  0.64] at (328.04, 28.36) {S6};

\node[text=drawColor,anchor=base,inner sep=0pt, outer sep=0pt, scale=  0.64] at (347.56, 28.36) {S7};

\node[text=drawColor,anchor=base,inner sep=0pt, outer sep=0pt, scale=  0.64] at (367.07, 28.36) {S8};
\end{scope}
\begin{scope}
\path[clip] (  0.00,  0.00) rectangle (507.79,142.26);
\definecolor{drawColor}{gray}{0.20}

\path[draw=drawColor,line width= 0.4pt,line join=round] (394.49, 34.37) --
	(394.49, 36.37);

\path[draw=drawColor,line width= 0.4pt,line join=round] (414.01, 34.37) --
	(414.01, 36.37);

\path[draw=drawColor,line width= 0.4pt,line join=round] (433.53, 34.37) --
	(433.53, 36.37);

\path[draw=drawColor,line width= 0.4pt,line join=round] (453.04, 34.37) --
	(453.04, 36.37);

\path[draw=drawColor,line width= 0.4pt,line join=round] (472.56, 34.37) --
	(472.56, 36.37);

\path[draw=drawColor,line width= 0.4pt,line join=round] (492.08, 34.37) --
	(492.08, 36.37);
\end{scope}
\begin{scope}
\path[clip] (  0.00,  0.00) rectangle (507.79,142.26);
\definecolor{drawColor}{gray}{0.30}

\node[text=drawColor,anchor=base,inner sep=0pt, outer sep=0pt, scale=  0.64] at (394.49, 28.36) {A1};

\node[text=drawColor,anchor=base,inner sep=0pt, outer sep=0pt, scale=  0.64] at (414.01, 28.36) {A2};

\node[text=drawColor,anchor=base,inner sep=0pt, outer sep=0pt, scale=  0.64] at (433.53, 28.36) {A3};

\node[text=drawColor,anchor=base,inner sep=0pt, outer sep=0pt, scale=  0.64] at (453.04, 28.36) {A4};

\node[text=drawColor,anchor=base,inner sep=0pt, outer sep=0pt, scale=  0.64] at (472.56, 28.36) {A5};

\node[text=drawColor,anchor=base,inner sep=0pt, outer sep=0pt, scale=  0.64] at (492.08, 28.36) {A6};
\end{scope}
\begin{scope}
\path[clip] (  0.00,  0.00) rectangle (507.79,142.26);
\definecolor{drawColor}{gray}{0.30}

\node[text=drawColor,anchor=base east,inner sep=0pt, outer sep=0pt, scale=  0.64] at ( 31.59, 34.21) {\SI{10}{ms}};

\node[text=drawColor,anchor=base east,inner sep=0pt, outer sep=0pt, scale=  0.64] at ( 31.59, 55.63) {\SI{100}{ms}};

\node[text=drawColor,anchor=base east,inner sep=0pt, outer sep=0pt, scale=  0.64] at ( 31.59, 77.05) {\SI{1}{s}};

\node[text=drawColor,anchor=base east,inner sep=0pt, outer sep=0pt, scale=  0.64] at ( 31.59, 98.47) {\SI{10}{s}};

\node[text=drawColor,anchor=base east,inner sep=0pt, outer sep=0pt, scale=  0.64] at ( 31.59,119.89) {\SI{100}{s}};
\end{scope}
\begin{scope}
\path[clip] (  0.00,  0.00) rectangle (507.79,142.26);
\definecolor{drawColor}{gray}{0.20}

\path[draw=drawColor,line width= 0.4pt,line join=round] ( 33.19, 36.41) --
	( 35.19, 36.41);

\path[draw=drawColor,line width= 0.4pt,line join=round] ( 33.19, 57.83) --
	( 35.19, 57.83);

\path[draw=drawColor,line width= 0.4pt,line join=round] ( 33.19, 79.25) --
	( 35.19, 79.25);

\path[draw=drawColor,line width= 0.4pt,line join=round] ( 33.19,100.67) --
	( 35.19,100.67);

\path[draw=drawColor,line width= 0.4pt,line join=round] ( 33.19,122.09) --
	( 35.19,122.09);
\end{scope}
\begin{scope}
\path[clip] (  0.00,  0.00) rectangle (507.79,142.26);
\definecolor{drawColor}{RGB}{0,0,0}

\node[text=drawColor,anchor=base,inner sep=0pt, outer sep=0pt, scale=  0.80] at (269.49,  5.56) {URL};
\end{scope}
\begin{scope}
\path[clip] (  0.00,  0.00) rectangle (507.79,142.26);
\definecolor{drawColor}{RGB}{0,0,0}

\node[text=drawColor,rotate= 90.00,anchor=base,inner sep=0pt, outer sep=0pt, scale=  0.80] at (  9.51, 79.65) {Median fetch time (log scale)};
\end{scope}
\begin{scope}
\path[clip] (  0.00,  0.00) rectangle (507.79,142.26);
\definecolor{fillColor}{RGB}{255,255,255}

\path[fill=fillColor] (107.55,130.92) rectangle (431.43,138.26);
\end{scope}
\begin{scope}
\path[clip] (  0.00,  0.00) rectangle (507.79,142.26);
\definecolor{fillColor}{RGB}{255,255,255}

\path[fill=fillColor] (111.55,123.81) rectangle (126.00,138.26);
\end{scope}
\begin{scope}
\path[clip] (  0.00,  0.00) rectangle (507.79,142.26);
\definecolor{fillColor}{RGB}{51,34,136}

\path[fill=fillColor] (118.77,131.04) circle (  1.96);
\end{scope}
\begin{scope}
\path[clip] (  0.00,  0.00) rectangle (507.79,142.26);
\definecolor{fillColor}{RGB}{255,255,255}

\path[fill=fillColor] (156.77,123.81) rectangle (171.22,138.26);
\end{scope}
\begin{scope}
\path[clip] (  0.00,  0.00) rectangle (507.79,142.26);
\definecolor{fillColor}{RGB}{204,102,119}

\path[fill=fillColor] (164.00,134.09) --
	(166.64,129.51) --
	(161.35,129.51) --
	cycle;
\end{scope}
\begin{scope}
\path[clip] (  0.00,  0.00) rectangle (507.79,142.26);
\definecolor{fillColor}{RGB}{255,255,255}

\path[fill=fillColor] (203.75,123.81) rectangle (218.21,138.26);
\end{scope}
\begin{scope}
\path[clip] (  0.00,  0.00) rectangle (507.79,142.26);
\definecolor{fillColor}{RGB}{17,119,51}

\path[fill=fillColor] (209.02,129.07) --
	(212.94,129.07) --
	(212.94,133.00) --
	(209.02,133.00) --
	cycle;
\end{scope}
\begin{scope}
\path[clip] (  0.00,  0.00) rectangle (507.79,142.26);
\definecolor{fillColor}{RGB}{255,255,255}

\path[fill=fillColor] (275.88,123.81) rectangle (290.34,138.26);
\end{scope}
\begin{scope}
\path[clip] (  0.00,  0.00) rectangle (507.79,142.26);
\definecolor{drawColor}{RGB}{170,68,153}

\path[draw=drawColor,line width= 0.4pt,line join=round,line cap=round] (280.33,131.04) -- (285.88,131.04);

\path[draw=drawColor,line width= 0.4pt,line join=round,line cap=round] (283.11,128.26) -- (283.11,133.81);
\end{scope}
\begin{scope}
\path[clip] (  0.00,  0.00) rectangle (507.79,142.26);
\definecolor{fillColor}{RGB}{255,255,255}

\path[fill=fillColor] (357.97,123.81) rectangle (372.42,138.26);
\end{scope}
\begin{scope}
\path[clip] (  0.00,  0.00) rectangle (507.79,142.26);
\definecolor{drawColor}{RGB}{17,119,51}

\path[draw=drawColor,line width= 0.4pt,line join=round,line cap=round] (363.23,129.07) rectangle (367.15,133.00);

\path[draw=drawColor,line width= 0.4pt,line join=round,line cap=round] (363.23,129.07) -- (367.15,133.00);

\path[draw=drawColor,line width= 0.4pt,line join=round,line cap=round] (363.23,133.00) -- (367.15,129.07);
\end{scope}
\begin{scope}
\path[clip] (  0.00,  0.00) rectangle (507.79,142.26);
\definecolor{drawColor}{RGB}{0,0,0}

\node[text=drawColor,anchor=base west,inner sep=0pt, outer sep=0pt, scale=  0.64] at (130.00,128.83) {Original};
\end{scope}
\begin{scope}
\path[clip] (  0.00,  0.00) rectangle (507.79,142.26);
\definecolor{drawColor}{RGB}{0,0,0}

\node[text=drawColor,anchor=base west,inner sep=0pt, outer sep=0pt, scale=  0.64] at (175.22,128.83) {Modified};
\end{scope}
\begin{scope}
\path[clip] (  0.00,  0.00) rectangle (507.79,142.26);
\definecolor{drawColor}{RGB}{0,0,0}

\node[text=drawColor,anchor=base west,inner sep=0pt, outer sep=0pt, scale=  0.64] at (222.21,128.83) {Blockaid (cached)};
\end{scope}
\begin{scope}
\path[clip] (  0.00,  0.00) rectangle (507.79,142.26);
\definecolor{drawColor}{RGB}{0,0,0}

\node[text=drawColor,anchor=base west,inner sep=0pt, outer sep=0pt, scale=  0.64] at (294.34,128.83) {Blockaid (cold cache)};
\end{scope}
\begin{scope}
\path[clip] (  0.00,  0.00) rectangle (507.79,142.26);
\definecolor{drawColor}{RGB}{0,0,0}

\node[text=drawColor,anchor=base west,inner sep=0pt, outer sep=0pt, scale=  0.64] at (376.42,128.83) {Blockaid (no cache)};
\end{scope}
\end{tikzpicture}

%% file: figures/winners.tex
\begin{tikzpicture}[x=1pt,y=1pt]
\definecolor{fillColor}{RGB}{255,255,255}
\path[use as bounding box,fill=fillColor,fill opacity=0.00] (0,0) rectangle (241.85,142.26);
\begin{scope}
\path[clip] (  0.00,  0.00) rectangle (241.85,142.26);
\definecolor{drawColor}{RGB}{255,255,255}
\definecolor{fillColor}{RGB}{255,255,255}

\path[draw=drawColor,line width= 0.4pt,line join=round,line cap=round,fill=fillColor] (  0.00,  0.00) rectangle (241.85,142.26);
\end{scope}
\begin{scope}
\path[clip] ( 28.04, 13.25) rectangle (107.51,119.30);
\definecolor{fillColor}{RGB}{255,255,255}

\path[fill=fillColor] ( 28.04, 13.25) rectangle (107.51,119.30);
\definecolor{drawColor}{gray}{0.92}

\path[draw=drawColor,line width= 0.2pt,line join=round] ( 28.04, 30.12) --
	(107.51, 30.12);

\path[draw=drawColor,line width= 0.2pt,line join=round] ( 28.04, 54.22) --
	(107.51, 54.22);

\path[draw=drawColor,line width= 0.2pt,line join=round] ( 28.04, 78.33) --
	(107.51, 78.33);

\path[draw=drawColor,line width= 0.2pt,line join=round] ( 28.04,102.43) --
	(107.51,102.43);

\path[draw=drawColor,line width= 0.4pt,line join=round] ( 28.04, 18.07) --
	(107.51, 18.07);

\path[draw=drawColor,line width= 0.4pt,line join=round] ( 28.04, 42.17) --
	(107.51, 42.17);

\path[draw=drawColor,line width= 0.4pt,line join=round] ( 28.04, 66.28) --
	(107.51, 66.28);

\path[draw=drawColor,line width= 0.4pt,line join=round] ( 28.04, 90.38) --
	(107.51, 90.38);

\path[draw=drawColor,line width= 0.4pt,line join=round] ( 28.04,114.48) --
	(107.51,114.48);

\path[draw=drawColor,line width= 0.4pt,line join=round] ( 42.94, 13.25) --
	( 42.94,119.30);

\path[draw=drawColor,line width= 0.4pt,line join=round] ( 67.78, 13.25) --
	( 67.78,119.30);

\path[draw=drawColor,line width= 0.4pt,line join=round] ( 92.61, 13.25) --
	( 92.61,119.30);
\definecolor{fillColor}{gray}{0.80}

\path[fill=fillColor] ( 31.77, 18.07) rectangle ( 54.12, 82.35);
\definecolor{fillColor}{gray}{0.20}

\path[fill=fillColor] ( 31.77, 82.35) rectangle ( 54.12,114.48);
\definecolor{fillColor}{gray}{0.80}

\path[fill=fillColor] ( 56.60, 18.07) rectangle ( 78.95, 95.53);
\definecolor{fillColor}{gray}{0.20}

\path[fill=fillColor] ( 56.60, 95.53) rectangle ( 78.95,114.48);
\definecolor{fillColor}{gray}{0.80}

\path[fill=fillColor] ( 81.44, 18.07) rectangle (103.79,101.36);
\definecolor{fillColor}{gray}{0.20}

\path[fill=fillColor] ( 81.44,101.36) rectangle (103.79,114.48);
\end{scope}
\begin{scope}
\path[clip] (111.51, 13.25) rectangle (190.99,119.30);
\definecolor{fillColor}{RGB}{255,255,255}

\path[fill=fillColor] (111.51, 13.25) rectangle (190.99,119.30);
\definecolor{drawColor}{gray}{0.92}

\path[draw=drawColor,line width= 0.2pt,line join=round] (111.51, 30.12) --
	(190.99, 30.12);

\path[draw=drawColor,line width= 0.2pt,line join=round] (111.51, 54.22) --
	(190.99, 54.22);

\path[draw=drawColor,line width= 0.2pt,line join=round] (111.51, 78.33) --
	(190.99, 78.33);

\path[draw=drawColor,line width= 0.2pt,line join=round] (111.51,102.43) --
	(190.99,102.43);

\path[draw=drawColor,line width= 0.4pt,line join=round] (111.51, 18.07) --
	(190.99, 18.07);

\path[draw=drawColor,line width= 0.4pt,line join=round] (111.51, 42.17) --
	(190.99, 42.17);

\path[draw=drawColor,line width= 0.4pt,line join=round] (111.51, 66.28) --
	(190.99, 66.28);

\path[draw=drawColor,line width= 0.4pt,line join=round] (111.51, 90.38) --
	(190.99, 90.38);

\path[draw=drawColor,line width= 0.4pt,line join=round] (111.51,114.48) --
	(190.99,114.48);

\path[draw=drawColor,line width= 0.4pt,line join=round] (126.42, 13.25) --
	(126.42,119.30);

\path[draw=drawColor,line width= 0.4pt,line join=round] (151.25, 13.25) --
	(151.25,119.30);

\path[draw=drawColor,line width= 0.4pt,line join=round] (176.09, 13.25) --
	(176.09,119.30);
\definecolor{fillColor}{gray}{0.80}

\path[fill=fillColor] (115.24, 18.07) rectangle (137.59, 91.26);
\definecolor{fillColor}{RGB}{152,152,152}

\path[fill=fillColor] (115.24, 91.26) rectangle (137.59,114.24);
\definecolor{fillColor}{gray}{0.20}

\path[fill=fillColor] (115.24,114.24) rectangle (137.59,114.48);
\definecolor{fillColor}{gray}{0.80}

\path[fill=fillColor] (140.08, 18.07) rectangle (162.43, 84.46);
\definecolor{fillColor}{RGB}{152,152,152}

\path[fill=fillColor] (140.08, 84.46) rectangle (162.43,109.50);
\definecolor{fillColor}{gray}{0.20}

\path[fill=fillColor] (140.08,109.50) rectangle (162.43,114.48);
\definecolor{fillColor}{gray}{0.80}

\path[fill=fillColor] (164.91, 18.07) rectangle (187.26, 57.98);
\definecolor{fillColor}{RGB}{152,152,152}

\path[fill=fillColor] (164.91, 57.98) rectangle (187.26,112.41);
\definecolor{fillColor}{gray}{0.20}

\path[fill=fillColor] (164.91,112.41) rectangle (187.26,114.48);
\end{scope}
\begin{scope}
\path[clip] ( 28.04,119.30) rectangle (107.51,138.26);
\definecolor{drawColor}{gray}{0.20}
\definecolor{fillColor}{gray}{0.85}

\path[draw=drawColor,line width= 0.4pt,line join=round,line cap=round,fill=fillColor] ( 28.04,119.30) rectangle (107.51,138.26);
\definecolor{drawColor}{gray}{0.10}

\node[text=drawColor,anchor=base,inner sep=0pt, outer sep=0pt, scale=  0.64] at ( 67.78,130.03) {\textbf{No cache}};

\node[text=drawColor,anchor=base,inner sep=0pt, outer sep=0pt, scale=  0.64] at ( 67.78,123.12) {(compliance checking only)};
\end{scope}
\begin{scope}
\path[clip] (111.51,119.30) rectangle (190.99,138.26);
\definecolor{drawColor}{gray}{0.20}
\definecolor{fillColor}{gray}{0.85}

\path[draw=drawColor,line width= 0.4pt,line join=round,line cap=round,fill=fillColor] (111.51,119.30) rectangle (190.99,138.26);
\definecolor{drawColor}{gray}{0.10}

\node[text=drawColor,anchor=base,inner sep=0pt, outer sep=0pt, scale=  0.64] at (151.25,130.03) {\textbf{Cache miss}};

\node[text=drawColor,anchor=base,inner sep=0pt, outer sep=0pt, scale=  0.64] at (151.25,123.12) {(template generation)};
\end{scope}
\begin{scope}
\path[clip] (  0.00,  0.00) rectangle (241.85,142.26);
\definecolor{drawColor}{RGB}{0,0,0}

\path[draw=drawColor,line width= 0.4pt,line join=round] ( 28.04, 13.25) --
	(107.51, 13.25);
\end{scope}
\begin{scope}
\path[clip] (  0.00,  0.00) rectangle (241.85,142.26);
\definecolor{drawColor}{gray}{0.20}

\path[draw=drawColor,line width= 0.4pt,line join=round] ( 42.94, 11.25) --
	( 42.94, 13.25);

\path[draw=drawColor,line width= 0.4pt,line join=round] ( 67.78, 11.25) --
	( 67.78, 13.25);

\path[draw=drawColor,line width= 0.4pt,line join=round] ( 92.61, 11.25) --
	( 92.61, 13.25);
\end{scope}
\begin{scope}
\path[clip] (  0.00,  0.00) rectangle (241.85,142.26);
\definecolor{drawColor}{gray}{0.30}

\node[text=drawColor,anchor=base,inner sep=0pt, outer sep=0pt, scale=  0.64] at ( 42.94,  5.24) {diaspora*};

\node[text=drawColor,anchor=base,inner sep=0pt, outer sep=0pt, scale=  0.64] at ( 67.78,  5.24) {Spree};

\node[text=drawColor,anchor=base,inner sep=0pt, outer sep=0pt, scale=  0.64] at ( 92.61,  5.24) {Autolab};
\end{scope}
\begin{scope}
\path[clip] (  0.00,  0.00) rectangle (241.85,142.26);
\definecolor{drawColor}{RGB}{0,0,0}

\path[draw=drawColor,line width= 0.4pt,line join=round] (111.51, 13.25) --
	(190.99, 13.25);
\end{scope}
\begin{scope}
\path[clip] (  0.00,  0.00) rectangle (241.85,142.26);
\definecolor{drawColor}{gray}{0.20}

\path[draw=drawColor,line width= 0.4pt,line join=round] (126.42, 11.25) --
	(126.42, 13.25);

\path[draw=drawColor,line width= 0.4pt,line join=round] (151.25, 11.25) --
	(151.25, 13.25);

\path[draw=drawColor,line width= 0.4pt,line join=round] (176.09, 11.25) --
	(176.09, 13.25);
\end{scope}
\begin{scope}
\path[clip] (  0.00,  0.00) rectangle (241.85,142.26);
\definecolor{drawColor}{gray}{0.30}

\node[text=drawColor,anchor=base,inner sep=0pt, outer sep=0pt, scale=  0.64] at (126.42,  5.24) {diaspora*};

\node[text=drawColor,anchor=base,inner sep=0pt, outer sep=0pt, scale=  0.64] at (151.25,  5.24) {Spree};

\node[text=drawColor,anchor=base,inner sep=0pt, outer sep=0pt, scale=  0.64] at (176.09,  5.24) {Autolab};
\end{scope}
\begin{scope}
\path[clip] (  0.00,  0.00) rectangle (241.85,142.26);
\definecolor{drawColor}{RGB}{0,0,0}

\path[draw=drawColor,line width= 0.4pt,line join=round] ( 28.04, 13.25) --
	( 28.04,119.30);
\end{scope}
\begin{scope}
\path[clip] (  0.00,  0.00) rectangle (241.85,142.26);
\definecolor{drawColor}{gray}{0.30}

\node[text=drawColor,anchor=base east,inner sep=0pt, outer sep=0pt, scale=  0.64] at ( 24.44, 15.87) {0.00};

\node[text=drawColor,anchor=base east,inner sep=0pt, outer sep=0pt, scale=  0.64] at ( 24.44, 39.97) {0.25};

\node[text=drawColor,anchor=base east,inner sep=0pt, outer sep=0pt, scale=  0.64] at ( 24.44, 64.07) {0.50};

\node[text=drawColor,anchor=base east,inner sep=0pt, outer sep=0pt, scale=  0.64] at ( 24.44, 88.17) {0.75};

\node[text=drawColor,anchor=base east,inner sep=0pt, outer sep=0pt, scale=  0.64] at ( 24.44,112.28) {1.00};
\end{scope}
\begin{scope}
\path[clip] (  0.00,  0.00) rectangle (241.85,142.26);
\definecolor{drawColor}{gray}{0.20}

\path[draw=drawColor,line width= 0.4pt,line join=round] ( 26.04, 18.07) --
	( 28.04, 18.07);

\path[draw=drawColor,line width= 0.4pt,line join=round] ( 26.04, 42.17) --
	( 28.04, 42.17);

\path[draw=drawColor,line width= 0.4pt,line join=round] ( 26.04, 66.28) --
	( 28.04, 66.28);

\path[draw=drawColor,line width= 0.4pt,line join=round] ( 26.04, 90.38) --
	( 28.04, 90.38);

\path[draw=drawColor,line width= 0.4pt,line join=round] ( 26.04,114.48) --
	( 28.04,114.48);
\end{scope}
\begin{scope}
\path[clip] (  0.00,  0.00) rectangle (241.85,142.26);
\definecolor{drawColor}{RGB}{0,0,0}

\node[text=drawColor,rotate= 90.00,anchor=base,inner sep=0pt, outer sep=0pt, scale=  0.80] at (  9.51, 66.28) {Fraction of wins};
\end{scope}
\begin{scope}
\path[clip] (  0.00,  0.00) rectangle (241.85,142.26);
\definecolor{fillColor}{RGB}{255,255,255}

\path[fill=fillColor] (198.99, 47.23) rectangle (237.85, 85.32);
\end{scope}
\begin{scope}
\path[clip] (  0.00,  0.00) rectangle (241.85,142.26);
\definecolor{drawColor}{RGB}{0,0,0}

\node[text=drawColor,anchor=base west,inner sep=0pt, outer sep=0pt, scale=  0.80] at (198.99, 79.04) {Solver};
\end{scope}
\begin{scope}
\path[clip] (  0.00,  0.00) rectangle (241.85,142.26);
\definecolor{fillColor}{RGB}{255,255,255}

\path[fill=fillColor] (198.99, 62.88) rectangle (210.37, 74.26);
\end{scope}
\begin{scope}
\path[clip] (  0.00,  0.00) rectangle (241.85,142.26);
\definecolor{fillColor}{gray}{0.20}

\path[fill=fillColor] (199.70, 63.59) rectangle (209.66, 73.55);
\end{scope}
\begin{scope}
\path[clip] (  0.00,  0.00) rectangle (241.85,142.26);
\definecolor{fillColor}{RGB}{255,255,255}

\path[fill=fillColor] (198.99, 51.50) rectangle (210.37, 62.88);
\end{scope}
\begin{scope}
\path[clip] (  0.00,  0.00) rectangle (241.85,142.26);
\definecolor{fillColor}{RGB}{152,152,152}

\path[fill=fillColor] (199.70, 52.21) rectangle (209.66, 62.17);
\end{scope}
\begin{scope}
\path[clip] (  0.00,  0.00) rectangle (241.85,142.26);
\definecolor{fillColor}{RGB}{255,255,255}

\path[fill=fillColor] (198.99, 40.12) rectangle (210.37, 51.50);
\end{scope}
\begin{scope}
\path[clip] (  0.00,  0.00) rectangle (241.85,142.26);
\definecolor{fillColor}{gray}{0.80}

\path[fill=fillColor] (199.70, 40.83) rectangle (209.66, 50.78);
\end{scope}
\begin{scope}
\path[clip] (  0.00,  0.00) rectangle (241.85,142.26);
\definecolor{drawColor}{RGB}{0,0,0}

\node[text=drawColor,anchor=base west,inner sep=0pt, outer sep=0pt, scale=  0.64] at (214.37, 66.36) {\textsc{cvc5}};
\end{scope}
\begin{scope}
\path[clip] (  0.00,  0.00) rectangle (241.85,142.26);
\definecolor{drawColor}{RGB}{0,0,0}

\node[text=drawColor,anchor=base west,inner sep=0pt, outer sep=0pt, scale=  0.64] at (214.37, 54.98) {Vampire};
\end{scope}
\begin{scope}
\path[clip] (  0.00,  0.00) rectangle (241.85,142.26);
\definecolor{drawColor}{RGB}{0,0,0}

\node[text=drawColor,anchor=base west,inner sep=0pt, outer sep=0pt, scale=  0.64] at (214.37, 43.60) {Z3};
\end{scope}
\end{tikzpicture}

%% file: discussions.tex
\paragraph{Comparison to row- and cell-level policy.}
Several commercial databases (such as SQL Server~\cite{msft21:sql_rls} and Oracle~\cite{oracle:vpd}) implement \emph{row- and/or cell-level} data-access policies,
which specify accessible information at the granularity of rows or cells.

Such policies are less expressive than the view-based ones supported by \name. 
For example, suppose we wish to allow each user to view everyone's timetables (i.e., the start and end times of the events they attend).
Querying someone's timetable requires joining the \textit{Events} and \textit{Attendances} tables on the \textit{EId} column, which must then be treated as visible by a cell-level policy.
But this inevitably reveals meeting attendee information as well.
Instead, we can implement this policy using a view:
\begin{lstlisting}
SELECT UId, StartTime, EndTime
FROM   Events e
JOIN   Attendances a ON e.EId = a.EId
\end{lstlisting}
which lists the times of events attended without revealing \textit{EId}.%

\paragraph{False rejections.}
\begin{mychange}%
Even though false rejections of compliant queries never occurred in our evaluation,
they remain a possibility for several reasons, including:
\begin{inparaenum}[(1)]
    \item approximate rewriting into basic queries, which is incomplete;
    \item our use of strong compliance; and
    \item solver timeouts.
\end{inparaenum}
Developers can reduce the chance of false rejections by running an application's end-to-end test suite under \name{} before deployment, and manually examining any rejected query to determine whether it is due to a false positive, a bug in the code, or a misspecified policy.

\end{mychange}

\paragraph{Off-path deployment.}
\begin{mychange}%
If an operator is especially worried about false rejections affecting a website's availability, we can modify \name{} to log potential violations instead of blocking any queries.
We can even move \name{} off-path by having the application stream its queries to \name{} to be checked asynchronously,
further reducing its performance impact.%
\end{mychange}

\paragraph{What if \name{} could issue its own queries?}
\begin{mychange}%
Suppose \name{} can issue extra queries---but only ones answerable using the views, lest the decision itself reveal sensitive data---when checking compliance.
\name{} can now safely allow more queries from the application.
For example, faced with the formerly non-compliant single query from \Cref{ex:noncompliant}:
\begin{lstlisting}
SELECT Title FROM Events WHERE EId = 5
\end{lstlisting}
\name{} can now \emph{ask} whether the user attends Event~\#5 and if so, allow the query.
In fact, under this setup the ``necessary-and-sufficient'' condition for application noninterference (in the sense of \Cref{sec:spec:ni}) becomes instance-based determinacy~\cite{rizvi04:rewriting,Koutris12:pricing,Zhang05:authorization-views},
a criterion less stringent than trace determinacy.

We decided against this design alternative for two reasons.
First, it seems nontrivial to check instance-based determinacy efficiently:
\name{} must either figure out a small set of queries to ask, a difficult problem, or fetch all accessible information, an expensive task.
Second, \name{} is designed for conventional applications that do not \emph{rely on} an enforcer for data-access compliance.
These applications should not be issuing queries that fail trace determinacy but pass instance-based determinacy: Such queries can, in \name's absence, reveal inaccessible information on \emph{another} database and typically indicate application bugs.
Thus, \name{} is right to flag them.%
\end{mychange}

\paragraph{Theoretically optimal templates.}
\begin{mychange}%
While decision templates produced by \name{} are general enough in practice,
they might not be \emph{maximally general} among all sound templates that match the query and trace being checked.
For one thing, the template condition might not be maximally weak (\Cref{sec:cache:core:cond}).
For another, a maximally general template can have a \emph{longer} trace than the concrete one, a possibility \name{} never explores.

Fundamentally, our template generation algorithm is limited by its \emph{black-box access} to the policy:
It interacts with the policy solely by checking template soundness using a solver.
Producing maximally general templates might require opening up this black box and having the policy guide template generation more directly,
a path we plan to explore in future work.%
\end{mychange}

%% file: conclusion.tex
\name enforces view-based data-access policies on web applications in a semantically transparent and backwards compatible manner.
It verifies policy compliance using SMT solvers and achieves low overhead using a novel caching and generalization technique.
We hope that \name's approach will help rule out data-access bugs in real-world applications.

%% file: artifactAppendix.tex
\section{Artifact Appendix}\label{sec:artifact}
\begin{table*}\small\centering
    \begin{threeparttable}[b]
        \caption{Where artifact contents are hosted.}\label{tbl:artifact}
        \renewcommand{\arraystretch}{1.1}
        \begin{tabular}{lll}
            \toprule
            \textbf{Content} & \textbf{Location} & \textbf{Branch / tag / release} \\ \midrule
            \textbf{Artifact README} & \url{https://github.com/blockaid-project/artifact-eval} & \texttt{main} branch \\
            \textbf{\name{} source} & \url{https://github.com/blockaid-project/blockaid} & \texttt{main} branch (latest version) \\
                                    &                                                    & \texttt{osdi22ae} branch (AE version)\tnote{a} \\
            \textbf{Experiment launcher} & \url{https://hub.docker.com/repository/docker/blockaid/ae} & \texttt{latest} tag \\
            \quad Launcher source & \url{https://github.com/blockaid-project/ae-launcher} & \texttt{main} branch \\
            \textbf{VM image} & \url{https://github.com/blockaid-project/ae-vm-image} & \texttt{osdi22ae} release \\
            \quad Experiment scripts & \url{https://github.com/blockaid-project/experiments} & \texttt{osdi22ae} branch \\
            \textbf{Applications} \\
            \quad \diaspora{} & \url{https://github.com/blockaid-project/diaspora} & \texttt{blockaid} branch\tnote{b} \\
            \quad Spree & \url{https://github.com/blockaid-project/spree} & \texttt{bv4.3.0-orig} branch (original)\tnote{c} \\
                        & & \texttt{bv4.3.0} branch (modified)\tnote{d} \\
            \quad Autolab & \url{https://github.com/blockaid-project/Autolab} & \texttt{bv2.7.0-orig} branch (original)\tnote{c} \\
                          & & \texttt{bv2.7.0} branch (modified)\tnote{d} \\
            \textbf{Policies for applications} & \url{https://github.com/blockaid-project/app-policies} & \texttt{main} branch \\
            \bottomrule
        \end{tabular}
        \begin{tablenotes}
          \item [a] The ``AE version'' is the version of \name{} used in artifact evaluation.
          \item [b] The same \diaspora{} branch is used for both baseline and \name{} measurements.  The code added for \name{} is gated behind conditionals that check whether \name{} is in use.
          \item [c] ``(original)'' denotes the original application modified only to run on top of JRuby.
          \item [d] ``(modified)'' denotes the ``(original)'' code additionally modified to work with \name{} (\Cref{sec:eval:change}).
        \end{tablenotes}
    \end{threeparttable}
\end{table*}

\subsection*{Abstract}
Our artifact includes our \name{} implementation, which is compatible with applications that can run atop the JVM and connect to a database via JDBC (\Cref{sec:impl}).
We also provide the three applications we used for our evaluation---modified according to \Cref{sec:eval:change}---as well as the data-access policy we wrote for each.
Finally, we provide a setup for reproducing the evaluation results from \Cref{sec:eval}.

\subsection*{Scope}
This artifact can be used to run the main experiments from this paper: the page load time (PLT) measurements (\Cref{sec:eval:plt}) and the fetch latency measurements (\Cref{sec:eval:breakdown} and \Cref{sec:eval:solvers}) on the three applications.
From these experiments, it generates \Cref{tbl:benchmark} (with URLs and descriptions omitted), \Cref{fig:fetch}, and \Cref{fig:solvers}.
Because the full experiment can be time- and resource-consuming (taking roughly 15~hours on six Amazon EC2 c4.8xlarge instances), the experiment launcher can be configured to take fewer measurement rounds at the expense of accuracy.

Our \name{} implementation can also be used to enforce data-access policies on new applications, as long as they have been modified to satisfy our requirements (\Cref{sec:design:code}), run atop the JVM, and connect to the database using JDBC (\Cref{sec:impl}).

\subsection*{Contents}

This artifact consists of our \name{} implementation, the three applications used in our evaluation (with modifications described in \Cref{sec:eval:change}), the data-access policy we wrote for each, and scripts and virtual machine image for running the experiments.

\subsection*{Hosting}

See \Cref{tbl:artifact}.

\subsection*{Requirements}
The experiment launcher, which relies on Docker, launches experiments on Amazon EC2 and so requires an AWS account.
By default, it uses six c4.8xlarge instances---to run the PLT and fetch latency experiments for the three applications simultaneously.
However, it can be configured to launch fewer instances at a time (e.g., to run the experiments serially, using one instance at a time).

%% file: extendedAppendix.tex
\section{Proof: From Query Compliance to Application Noninterference}\label{sec:proofs}
\ifextended{}
\begin{figure*}
    \centering
    \begin{subfigure}[b]{.3\textwidth}
        \begin{align*}
            V^{\ctx}(D_1) &\diffBox{$=$} V^{\ctx}(D_2) \tikzmark{eq3L} \\
            Q_i(D_1) &\diffBox{$=$} O_i \tikzmark{eq1L} \\
            Q_i(D_2) &\diffBox{$=$} O_i \tikzmark{eq2L} \\
            \midrule
            Q(D_1) &\diffBox{$=$} Q(D_2) \tikzmark{conclusionL}
        \end{align*}
        \vspace{-2em}
        \caption{Compliance (\Cref{def:compliance}).}
    \end{subfigure}\hspace{.2\textwidth}%
    \begin{subfigure}[b]{.4\textwidth}
        \begin{align*}
            \tikzmark{eq3R} V^{\ctx}(D_1) &\diffBox{\textcolor{red}{$\subseteq$}} V^{\ctx}(D_2) & (\forall V\in\V) \\
            \tikzmark{eq1R} Q_i(D_1) &\diffBox{\textcolor{red}{$\supseteq$}} O_i & (\forall 1\leq i\leq n) \\
            \tikzmark{eq2R} Q_i(D_2) &\diffBox{\textcolor{red}{$\supseteq$}} O_i \tikzmark{eq2RR} & (\forall 1\leq i\leq n) \\
            \midrule
            \tikzmark{conclusionR} Q(D_1) &\diffBox{\textcolor{red}{$\subseteq$}} Q(D_2)
        \end{align*}
        \vspace{-2em}
        \caption{Strong compliance (\Cref{def:strong-compliance}).}
    \end{subfigure}
    \caption{The definition of compliance is turned into that of strong compliance in five steps.
    Dashed arrows for Steps~1--4 denote modifications; the solid line (strikethrough) for Step~5 denotes removal.}\label{fig:complianceMod}

    \newcommand*\circled[1]{\tikz[baseline=(char.base)]{
        \node[shape=circle,inner sep=1.5pt,fill=red,text=white] (char) {#1};}}
    \begin{tikzpicture}[overlay,remember picture,stepArrow/.style={-Latex,red,thick,dashed}]
        \draw[stepArrow] ([shift={(2ex,.5ex)}]pic cs:conclusionL) -- node {\circled{1}} ([shift={(-2ex,.5ex)}]pic cs:conclusionR);
        \draw[stepArrow] ([shift={(2ex,.5ex)}]pic cs:eq1L) -- node {\circled{2}} ([shift={(-2ex,.5ex)}]pic cs:eq1R);
        \draw[stepArrow] ([shift={(2ex,.5ex)}]pic cs:eq2L) -- node {\circled{3}} ([shift={(-2ex,.5ex)}]pic cs:eq2R);
        \draw[stepArrow] ([shift={(2ex,.5ex)}]pic cs:eq3L) -- node {\circled{4}} ([shift={(-2ex,.5ex)}]pic cs:eq3R);
        \draw[red,thick] ([shift={(-.5ex,.5ex)}]pic cs:eq2R) -- ([shift={(.5ex,.5ex)}]pic cs:eq2RR);
        \node at ([shift={(3ex,.5ex)}] pic cs:eq2RR) {\circled{5}};
    \end{tikzpicture}
\end{figure*}
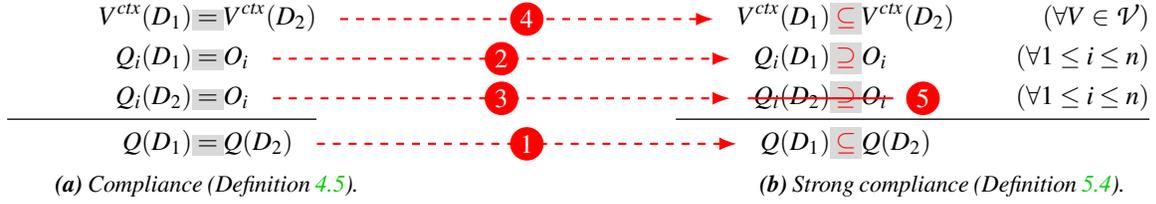

\begin{proof}[Proof of \Cref{thm:compliance}]
\begin{mychange}%
    We prove each part separately:

    \paragraph{Part~1.}
    Suppose $\EnforcePred(\ctx,Q,\T)=\cmark$ only when $Q$ is $\ctx$-compliant to~$\V$ given~$\T$.
    Pick any $\Prog$, $\ctx$, and $\req$, and let $D_1$ and $D_2$ be databases such that $V^{\ctx}(D_1)=V^{\ctx}(D_2)$ for all $V\in\V$.

    Consider executions $\Prog^{\EnforcePred}(\ctx,\req,D_1)$ and $\Prog^{\EnforcePred}(\ctx,\req,D_2)$.
    We will show that the two executions coincide, by induction on the number of steps taken by~$\Prog$.
    This will imply that $\Prog^{\EnforcePred}(\ctx,\req,D_1)=\Prog^{\EnforcePred}(\ctx,\req,D_2)$, finishing the proof.
    \begin{description}
        \item[Base case] Because $\Prog$ is assumed to be deterministic, so is $\Prog^{\EnforcePred}$, and so the two executions start off with the same program state.
        \item[Inductive step] Suppose the two executions coincide after $\Prog$ has taken $i$ steps.
            Consider the $i+1$st step taken on both sides:
            \begin{itemize}
                \item Suppose this step is a query $Q$ to the database.
                    Let $\T$ denote the (same) trace maintained by the two executions so far.
                    If $\EnforcePred(\ctx,Q,\T)=\xmark$, then both executions terminate with an error.
                    Otherwise, $Q$ must be $\ctx$-compliant to~$\V$ given~$\T$.
                    By assumption, $V^{\ctx}(D_1)=V^{\ctx}(D_2)$ for all $V\in\V$;
                    and by the construction of $\Prog^{\EnforcePred}$,
                    $Q_i(D_1)=Q_i(D_2)=O_i$ for every $(Q_i,O_i)\in\T$.
                    Therefore, we must have $Q(D_1)=Q(D_2)$, and so the two executions end up in the same program state after this step.
                \item If this step is \emph{not} a database query, then its behavior depends only on $\Prog$'s program state and inputs $\ctx$ and $\req$,
                    all of which are the same across the two executions at this time.
            \end{itemize}
    \end{description}

    \paragraph{Part~2.}
    Suppose that $\EnforcePred$ correctly enforces~$\V$.
    Pick any $\ctx$, $Q$, and prefix $\EnforcePred$-allowed $\T=\{(Q_i,O_i)\}_{i=1}^{n}$ such that $\EnforcePred(\ctx,Q,\T)=\cmark$.
    Consider the following program~$\Prog$:
    \begin{algorithmic}[lines=0]
        \Procedure{$\Prog$}{$\ctx,\req,D$}
        \For{$i\gets 1..n$}
            \State issue $Q_i(D)$ and discard the result
        \EndFor
        \State \textbf{return} $Q(D)$
        \EndProcedure
    \end{algorithmic}

    To show that $Q$ is $\ctx$-compliant to~$\V$ given~$\T$, let $D_1$ and $D_2$ be databases such that:
    \begin{align}
        V^{\ctx}(D_1) &= V^{\ctx}(D_2), & (\forall V\in\V) \label{eqn:appendixB:V} \\
        Q_i(D_1) &= O_i, & (\forall 1\leq i\leq n) \\
        Q_i(D_2) &= O_i. & (\forall 1\leq i\leq n)
    \end{align}
    Let $\req$ be a request, and
    consider executions $\Prog^{\EnforcePred}(\ctx,\req,D_1)$ and $\Prog^{\EnforcePred}(\ctx,\req,D_2)$.
    Because $\T$ is prefix $\EnforcePred$-allowed, neither execution ends in a policy violation error.
    Therefore,
    \begin{align*}
        \Prog^{\EnforcePred}(\ctx,\req,D_1) &= \Prog(\ctx,\req,D_1) = Q(D_1), \\
        \Prog^{\EnforcePred}(\ctx,\req,D_2) &= \Prog(\ctx,\req,D_2) = Q(D_2).
    \end{align*}
    Furthermore, because $\EnforcePred$ correctly enforces~$\V$, \Cref{eqn:appendixB:V} implies that
    $\Prog^{\EnforcePred}(\ctx,\req,D_1)=\Prog^{\EnforcePred}(\ctx,\req,D_2)$.
    We thus have $Q(D_1)=Q(D_2)$, concluding $Q$ to be $\ctx$-compliant to~$\V$ given~$\T$.
\end{mychange}
\end{proof}
\fi

\section{From Compliance to Strong Compliance}\label{sec:strong}
\ifextended
\begin{mychange}%
    To understand when strong compliance fails to coincide with compliance,
    let us look at \Cref{fig:complianceMod}, which illustrates how we modified the definition of compliance (\Cref{def:compliance}) into that of strong compliance (\Cref{def:strong-compliance}) in five steps.

    Step~1 does not affect the truthfulness of the formula since $D_1$ and $D_2$ are symmetric.
    Steps~2--3 adopt an \emph{open-world assumption} (OWA)~\cite{reiter77:closed},
    treating every query as returning partial results.
    Under this assumption, a trace can no longer represent the \emph{nonexistence} of a returned row;
    this can cause \name{} may falsely reject a query.
    However, such cases never arose in our evaluation.
    The OWA also proves convenient during decision template generation (\Cref{sec:cache:core:min})
    when \name{} computes a minimal sub-trace (which, by necessity, represents partial information) that guarantees strong compliance.

    To see how Step~4 affects the definition,
    suppose there are no database dependencies and the trace is empty (so Steps~2--3 are irrelevant).
    In this scenario, compliance holds iff $\V$ \emph{determines} $Q$~\cite{Nash10:determinacy,Segoufin05:determinacy},
    while strong compliance states that $Q$ has a \emph{monotonic} rewriting using $\V$.
    There are cases where determinacy holds but no monotonic rewriting exists;
    e.g., Nash~et~al.~\cite[\S~5.1]{Nash10:determinacy} present an example in terms of conjunctive queries.

    Finally, in Step~5 we drop the condition that $D_2$ be consistent with the trace.
    We can show by induction that this condition is redundant as long as each query in the trace is strongly compliant given the trace before it
    (which is the case in \name).
\end{mychange}
\fi